\documentclass[]{aastex631}

\shorttitle{Tracking Accretion in TNG50}
\shortauthors{Forbes et al.}

\graphicspath{{./}{figures/}}

\begin{document}

\title{Gas Accretion Can Drive Turbulence in Galaxies}

\author[0000-0002-1975-4449]{John C. Forbes}
\affiliation{Center for Computational Astrophysics, Flatiron Institute, 162 5th Avenue, New York, NY 10010, USA}

\correspondingauthor{John C. Forbes}
\email{jforbes@flatironinstitute.org}

\author[0000-0002-2791-5011]{Razieh Emami}\affiliation{Center for Astrophysics $|$ Harvard \& Smithsonian, 60 Garden St., Cambridge, MA 02138, USA}

\author{Rachel S. Somerville}
\affiliation{Center for Computational Astrophysics, Flatiron Institute, 162 5th Avenue, New York, NY 10010, USA}

\author{Shy Genel}
\affiliation{Center for Computational Astrophysics, Flatiron Institute, 162 5th Avenue, New York, NY 10010, USA}
\affiliation{Columbia Astrophysics Laboratory, Columbia University, New York, NY, 10027, USA}

\author{Dylan Nelson}
\affiliation{Zentrum für Astronomie der Universität Heidelberg, ITA, Albert-Ueberle-Str. 2, D-69120 Heidelberg, Germany}

\author[0000-0003-1065-9274]{Annalisa Pillepich}
\affiliation{Max-Planck-Institut fur Astronomie, Königstuhl 17, 69117 Heidelberg, Germany}

\author{Blakesley Burkhart}
\affiliation{Department of Physics and Astronomy, Rutgers University, 136 Frelinghuysen Road, Piscataway, NJ 08854, USA}
\affiliation{Center for Computational Astrophysics, Flatiron Institute, 162 5th Avenue, New York, NY 10010, USA}

\author[0000-0003-2630-9228]{Greg L. Bryan}
\affiliation{Department of Astronomy, Columbia University, 550 W 120th Street, New York, NY 10027, USA}
\affiliation{Center for Computational Astrophysics, Flatiron Institute, 162 5th Avenue, New York, NY 10010, USA}

\author[0000-0003-3893-854X]{Mark R. Krumholz}
\affiliation{Research School of Astronomy and Astrophysics, Australian National University, Canberra, ACT 2611 Australia}
\affiliation{ARC Centre of Excellence for Astronomy in Three Dimensions (ASTRO-3D), Canberra, ACT 2611 Australia}

\author{Lars Hernquist}
\affiliation{Center for Astrophysics $|$ Harvard \& Smithsonian, 60 Garden St., Cambridge, MA 02138, USA}

\author{Stephanie Tonnesen}
\affiliation{Center for Computational Astrophysics, Flatiron Institute, 162 5th Avenue, New York, NY 10010, USA}

\author{Paul Torrey}
\affiliation{Department of Astronomy, University of Florida, Gainesville, FL, USA}

\author{Viraj Pandya}
\affiliation{Department of Astronomy, Columbia University, 550 W 120th Street, New York, NY 10027, USA}

\author[0000-0003-4073-3236]{Christopher C. Hayward}
\affiliation{Center for Computational Astrophysics, Flatiron Institute, 162 5th Avenue, New York, NY 10010, USA}

\begin{abstract}

The driving of turbulence in galaxies is deeply connected with the physics of feedback, star formation, outflows, accretion, and radial transport in disks. The velocity dispersion of gas in galaxies therefore offers a promising observational window into these processes. However, the relative importance of each of these mechanisms remains controversial. In this work we revisit the possibility that turbulence on galactic scales is driven by the direct impact of accreting gaseous material on the disk. We measure this effect in a disk-like star-forming galaxy in IllustrisTNG, using the high-resolution cosmological magnetohydrodynamical simulation TNG50. We employ Lagrangian tracer particles with a high time cadence of only a few Myr to identify accretion and other events, such as star formation, outflows, and movement within the disk. The energies of particles as they arrive in the disk are measured by stacking the events in bins of time before and after the event. The average effect of each event is measured on the galaxy by fitting explicit models for the kinetic and turbulent energies as a function of time in the disk. These measurements are corroborated by measuring the cross-correlation of the turbulent energy in the different annuli of the disk with other time series, and searching for signals of causality, i.e. asymmetries in the cross-correlation across zero time lag. We find that accretion contributes to the large-scale turbulent kinetic energy even if it is not the dominant driver of turbulence in this $\sim 5 \times 10^{9} M_\odot$ stellar mass galaxy. Extrapolating this finding to a range of galaxy masses, we find that there are regimes where energy from direct accretion may dominate the turbulent energy budget, particularly in disk outskirts, galaxies less massive than the Milky Way, and at redshift $\sim 2$.

\end{abstract}

\section{Introduction} \label{sec:intro}

The velocity dispersion of gas in galactic disks is observable across cosmic time \citep[e.g.][]{forsterschreiber_sins_2009, genzel_sins_2014, wisnioski_kmos3d_2015, simons_epoch_2017, ubler_evolution_2019}, and provides an indicator of the physics at play in the regulation of the properties of the disk \citep{burkhart2021}. Galactic velocity dispersions in excess of the sound speed of the Warm Neutral Medium are often considered to be an indicator of turbulence on the scale of a disk scale height, though this can only be studied in detail with high resolution maps in nearby galaxies \citep{elmegreen_fractal_2001, sharda_first_2022}, and near this limit the measurements are likely to be sensitive to the procedure used to subtract off the thermal component and the expansion of the HII regions whose emission is being measured \citep[e.g.][]{law_sdssiv_2021}. Turbulence decays on a dynamical time \citep{maclowKineticEnergyDecay1998, stoneDissipationCompressibleMagnetohydrodynamic1998}, so large velocity dispersions must be continuously driven by some source or sources. The primary driver of turbulence on galactic scales has variously been argued to be star formation feedback \citep{joung_dependence_2009, green_high_2010, orr_swirls_2020a, law_sdssiv_2021}, Toomre \citep{toomre_gravitational_1964} disk instability \citep{dekel_formation_2009, krumholz_dynamics_2010, genzel_sins_2011, forbes_evolving_2012, cacciato_evolution_2012, forbes_balance_2014}, or the direct effects of accretion onto the disk\footnote{Some authors conflate the effects of accretion and inflows within the disk, since the energy that drives the turbulence in the latter case is, ultimately, the average inward motion within the disk of gas down the potential well of the galaxy. In this work, accretion-driven turbulence and in-disk inflow-driven turbulence refer to separate processes.} \citep{klessen_accretiondriven_2010,genel_effect_2012,gabor_delayed_2014}.  Observationally, the velocity dispersion of a galaxy is tied to the galaxy's mass, redshift, and star formation rate \citep{green_high_2010, kassin_epoch_2012, genzel_sins_2014, simons_transition_2015, simons_kinematic_2016, krumholz_turbulence_2016, krumholz_unified_2018}, though it is unclear which, if any, of these is the ``primary" correlate; efforts to disentangle these in nearby galaxies, where ionized gas, neutral gas, and stellar data are all available, remain a work in progress \citep[e.g.,][]{fisher_testing_2019, yu_major_2021}. In general, especially at high redshift, it is challenging to confidently extract velocity dispersions from integral field unit or slit-based spectra owing to a convolution of the rotational velocity of the galaxy within the spatial resolution of the observation \citep{davies_how_2011}, leading to the possibility favored in \cite{forbes_radially_2019} that high velocity dispersions at high redshift may be the result of systematic errors.

A key barrier to understanding the physical processes that drive the velocity dispersion is that all of the relevant quantities tend to be correlated with each other: velocity dispersion, star formation rate, gas surface density, and accretion rate. This is likely not a coincidence, since the depletion time, namely the gas mass divided by the star formation rate, is short compared to cosmological timescales for essentially all star-forming galaxies \citep{genzel_study_2010, bigiel_constant_2011, rodighiero_lesser_2011, saintonge_validation_2013, tacconi_phibss_2013, genzel_combined_2015}. Even in dwarfs the outflow rate may be large enough that the loss timescale, namely the gas mass divided by the mass loss rate (star formation rate plus galactic wind outflow rate) may be comparably short \citep{forbes_origin_2014}. This means that not only are galaxies globally in an approximate equilibrium state, at least as far as their accretion rate, gas content, star formation rate, dust content, and metallicity are concerned \citep{bouche_impact_2010, dave_analytic_2012, lilly_gas_2013, forbes_origin_2014, feldmann_equilibrium_2015, rodriguez-puebla_mainsequence_2016}, but deviations from this equilibrium, e.g. as the result of variable accretion, themselves follow well-defined patterns \citep{forbes_origin_2014, genzel_combined_2015, tacchella_confinement_2016, tacchella_stochastic_2020}. Even within galaxies, the column density and especially the velocity dispersion are expected to maintain profiles close to equilibrium \citep{forbes_balance_2014}. Under these conditions, correlations between any two quantities cannot necessarily be interpreted causally.

A particularly important example of this is the correlation between star formation rate and gas velocity dispersion, which has been interpreted as evidence for turbulence driven by stellar feedback \citep{green_high_2010}. Within the context of any particular model, the danger of interpreting this correlation as causation may be verified directly. \citet{krumholz_turbulence_2016} and \citet{krumholz_unified_2018} showed that for disks where the velocity dispersion is explicitly set by a balance between turbulent dissipation and either momentum injection from supernovae or Toomre instability, galaxies with higher star formation rates will generically have higher velocity dispersions in both cases. A qualitative comparison with the data suggests the feedback from star formation is dominant when $\mathrm{SFR} \lesssim 1 M_\odot/\mathrm{yr}$, while gravitational instability is necessary to produce large velocity dispersions, in excess of $\sim 20$ km/s. Shearing box experiments \citep{colling_impact_2018} find that the flow of mass through a disk plays an important role in its energy balance: disks with shear, where inflow releases energy, require considerably less star formation feedback to maintain the same velocity dispersion. Similarly, \citet{brucy_largescale_2020} find that gas-rich disks with properties chosen to mimic those found in giant disks at $z\sim 2$ lie on the observed Kennicutt-Schmidt relation only if gravitational instability provides more energy input to turbulence than star formation feedback; this is consistent with the earlier findings of \citet{joung_dependence_2009}, who show that, in simulations that impose by fiat a supernova rate consistent with the observed Kennicutt-Schmidt relation, and include no other source of energy input, gas-rich disks do not reach the velocity dispersions $\gtrsim 10$ km/s required by observations.

Another source of turbulence, namely the energy imparted by the direct accretion of material from the CGM to the disk, has been proposed in the past \citep{klessen_accretiondriven_2010,genel_effect_2012,gabor_delayed_2014}. The energy budget available for heating the disk from direct accretion is, per unit mass accreted, $\sim \epsilon v_\mathrm{circ}^2$, where $\epsilon$ is an efficiency factor $<1$ that \citet{klessen_accretiondriven_2010} argue is of order the density contrast between the accreting material and the denser material onto which it is accreting. Meanwhile, for supernovae, the energy budget per unit mass of stars formed available to drive turbulence and galactic winds \citep{hayward_how_2017} is of order $\langle p_*/m_* \rangle \sigma$ \citep[see the discussion in][]{krumholz_unified_2018}. The first factor is the average momentum injected by supernovae per mass of stars formed, which has a canonical value of 3000 km s$^{-1}$ \citep[e.g.][]{cioffi_dynamics_1988, thornton_energy_1998, kim_momentum_2015,  walch_energy_2015, gentry_enhanced_2017}, and $\sigma$ is the velocity dispersion of the gas itself. It follows that even for relatively large values of $v_\mathrm{circ}/\sigma$, say $\approx 10$, and even for massive galaxies with $v_\mathrm{circ}$ of a few hundred km/s, supernovae will dominate the energy budget by about a factor of $1/\epsilon$ provided that the accretion rate and the star formation rate are comparable locally. While this may be a reasonable first approximation, there are several important cases where this will not be the case. First, in galaxies where the mass loading factor $\eta$, i.e. the outflow rate per unit star formation rate, far exceeds unity, the global mass budget of the galaxy in a self-regulated equilibrium state is such that the accretion rate will be larger than the star formation rate by a factor of $\eta$. This discrepancy can be even larger if the galaxy is small enough that the depletion time is long, e.g. if the star formation rate is suppressed by a lack of molecular hydrogen \citep{krumholz_metallicitydependent_2012, krumholz_star_2013}, or more accurately a lack of the cold gas that is often correlated with molecular hydrogen \citep{krumholz_which_2011, krumholz_star_2012, forbes_suppression_2016}, the cosmological UV background \citep{okamoto_mass_2008}, or some other form of preventative feedback \citep{pandya_first_2020, 2022ApJ...927...75A}. Similarly, in the outskirts of galaxies where the gas surface density is low and dominated by atomic or ionized hydrogen, the star formation rate may be low, but substantial accreting material may still arrive there \citep[e.g.][]{stevens_how_2017, forbes_radially_2019, trapp_gas_2022, hafen_hotmode_2022}.

Numerical evidence against the importance of accreting material has been presented in the context of idealized simulations from \citet{hopkins_accretion_2013}, who showed that an isolated galaxy being fed by a well-defined cosmological stream had no discernible effect on the galaxy's velocity dispersion. Work employing fully cosmological zoom-in simulations at $z=0$ with a similar feedback prescription \citep{orr_swirls_2020a} came to a similar conclusion, although caution is required because these authors group accretion and inflow into the same category. This grouping is not unreasonable given that the energy in both cases must ultimately arise from gas flowing down a gravitational potential well, but in this work we separate them into two distinct terms in the energy equation. The direct effects of the accretion flow itself as it joins the disk, i.e. the exact process studied by \citet{hopkins_accretion_2013}, is our primary focus in this work. We will argue that accretion may in fact play an important role in setting the velocity dispersion, and in so doing we will provide an explanation for why we come to the opposite conclusion as \citet{hopkins_accretion_2013}. In short, their experiments do not cover the regime where accretion is likely to be important, and, in contrast to \citet{hopkins_accretion_2013}, the techniques we employ are able to detect and characterize the energy input from each term in the energy equation even when they are not the dominant source of turbulent energy, at the cost of a great deal of complexity in the analysis.

In this work, we characterize the direct effects of accretion on turbulence in the disk of a single IllustrisTNG galaxy with a $z=0$ stellar mass of $5 \times 10^9 M_\odot$. We combine two key pieces of technology: the high-cadence output in a sub-region of the full simulation volume, and the Lagrangian tracer particles that follow the motion of gas and baryonic collisionless particles (i.e. stars and outflow particles) in the simulation. This allows us to pinpoint accretion ``events'' and their local effects on the disk, and account for essentially every bit of mass that flows in or out of different regions of the galaxy. In section \ref{sec:tng} we review the relevant details of IllustrisTNG and TNG50 in particular. We then detail the selection and properties of the galaxy that we study in Section \ref{sec:galaxy}, along with the tracer particle events that we track. In Section \ref{sec:res}, we carry out four analyses to understand the effects of accretion on the disk. First, we stack the properties of the tracer particles before, during, and after an event, where we observe a distinct drop in the specific energy of accreting particles that we interpret as energy being shared with the disk. Second, we construct a regression model where we assume that the kinetic energy in every annulus in the disk as a function of time can be described as the sum of energies associated with each type of tracer-based event, in addition to some local heating and cooling terms. Third, we cross-correlate the specific energy in each annulus with several time series (of physically motivated quantities) and search for asymmetries in the cross-correlation functions across zero time lag, indicative of a causal relationship. Fourth, we compute another regression model, this time for the turbulent kinetic energy only -- this model is ultimately closest to fitting the exact quantities we are interested in, but the other analyses provide useful corroborating evidence for the direct effects of accretion. In Section \ref{sec:disc} we explore the implications of the accretion energy that we measured in the prior section. In particular, we conservatively extrapolate our results to other regimes beyond the single galaxy for which we have measured these quantities, and offer an explanation for the disagreements in the literature about which processes drive turbulence in galaxies. Finally we conclude in Section \ref{sec:conclusion}.

\section{TNG50} \label{sec:tng}

IllustrisTNG \citep{springel_first_2018,pillepich_first_2018, marinacci_first_2018,naiman_first_2018,nelson_first_2018} is a set of big-box cosmological simulations based on the earlier Illustris simulation \citep{vogelsberger_introducing_2014,vogelsberger_properties_2014,genel_introducing_2014}. IllustrisTNG was designed to produce a realistic population of galaxies from well-defined cosmological initial conditions through the evolution of the equations of magnetohydrodynamics, self-gravity, and star and black hole formation, feedback, and dynamics, much of which is accomplished with subgrid recipes \citep{weinberger_simulating_2017,pillepich_simulating_2018}. Of this set of simulations, TNG50 is the smallest in volume (51.7$^3$ comoving Mpc$^3$), but the highest in resolution, with baryonic and dark matter particles of masses $8.5 \times 10^4 M_\odot$ and $4.5 \times 10^5 M_\odot$ respectively \citep{pillepich_first_2019,nelson_first_2019,nelson_illustristng_2019}. 

IllustrisTNG and its predecessors have relied on modified versions of the \citet{springel_cosmological_2003} ISM model, which modifies the equation of state of the gas in galaxies at densities above the star formation threshold of 0.13 cm$^{-3}$. In particular, higher density gas is taken to have higher internal energy and hence pressure, the result of a self-regulated two-phase equilibrium between cold star-forming gas and hot volume-filling supernova-heated gas. Because the recorded internal energy of this phase is determined directly by the density in the context of this sub-grid model, we will instead focus only on the reasonably well-resolved components of the energy, namely the gravitational potential energy and the kinetic energy associated with the mean and dispersion of different velocity components within sub-regions of the disk. The evolution of gas velocity dispersions in TNG has been studied extensively \citep{pillepich_first_2019, ubler_kinematics_2021}, demonstrating good agreement between observations \citep[e.g.][]{wisnioski_kmos3d_2015, simons_epoch_2017, swinbank_angular_2017} and the trends seen in the simulations. The scalings derived by \citet{pillepich_first_2019} have even been used to substitute for unavailable observations in analytic models \citep{sharda_origin_2021, sharda_physics_2021, sharda_role_2021}.

Within the full periodic box are several ``subboxes'', which are sub-volumes of the full simulation for which data were recorded more frequently, i.e. about 3600 snapshots compared to 100 for the full simulation, though of course each samples a much more limited environment than the full box. The galaxy that we follow in this work is contained in TNG50 Subbox-0, which is $4 h^{-1}$ comoving Mpc on each side. The extremely fine time resolution, on average 4 Myr, allows us to pinpoint ``events'' and understand the state of the galaxy, on average, before and after such events.

Just as in the full snapshots, the subbox snapshots contain information about the Lagrangian tracer particles \citep{genel_following_2013}. Within the simulations, the tracers follow the motion of the fluid by, at any given moment, being associated with a particular gas cell or baryonic particle, and probabilistically being transferred to another computational element if there is some transfer of mass between the elements, including splitting or merging of cells, fluxes from Riemann problems, star formation, and feedback processes \citep{nelson_moving_2013}. The simulation is initialized with 1 tracer particle per baryonic mass element, so each tracer particle corresponds to $8.5 \times 10^4 M_\odot$ of baryons.

\section{The galaxy} \label{sec:galaxy}

\begin{figure}
    \centering
    \includegraphics[width=1\textwidth]{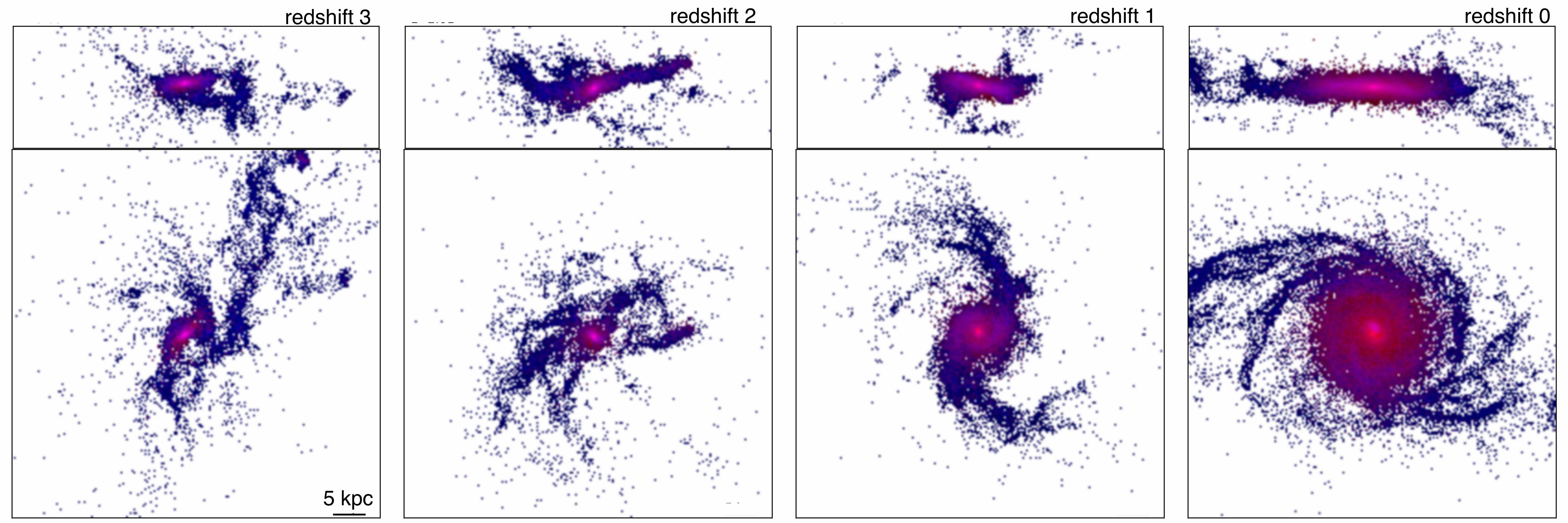}
    \caption{Images of the galaxy at different redshifts. Top panels are 60 kpc x 20 kpc edge-on views, and bottom panels are 60 kpc x 60 kpc face-on views with respect to the smoothed angular momentum direction at that moment. Here and throughout kpc are physical, not comoving. Tracer particles associated with stars are shown in red, and gas is shown in blue, with identical logarithmic stretches and a pixel width of 300 pc. Only bins with 3 or more particles are shown, meaning that the minimum column density is of order 3 $M_\odot\ {\rm pc}^{-2}$. The depth of the projection is $\pm 10$ kpc from the $x=0$ or $z=0$ plane in the top and bottom panels respectively.}
    \label{fig:postage}
\end{figure}

In order to study the evolution of a particular galaxy, we first identify all dark matter halos and subhalos in one of the subboxes using the SUBFIND catalogues available for the full simulation at $z=0$. The precomputed catalogues include basic properties of the galaxy, such as its position and total mass at the times of the full-box snapshots. We identify the most massive galaxies in the subbox, and search for those that are farthest from the subbox boundary. This is to prevent, to the degree possible, the inclusion of tracer particles that arrive in the galaxy from outside the subbox, whose data are not available at high time cadence. In Subbox0 there are 9 galaxies whose (sub)halo masses exceed $5 \times 10^{10} M_\odot$ at $z=0$ and which were within the subbox at $z=4$ as well. Of these, 4 galaxies are within the subbox at all times. One of these galaxies is nonetheless frequently within 100 kpc of the edge of the subbox. Of the 3 remaining galaxies, by far the most massive has a halo mass of $4.8 \times 10^{11} M_\odot$ at $z=0$. In the full box snapshot at $z=0$, this galaxy corresponds to subhalo 585079. While there are two other subboxes in the simulation (which probe different environments), and strictly speaking it is not necessary for the galaxy to {\it always} be in the subbox for this analysis, this galaxy certainly represents the most straightforward target to analyze. Rather than attempt to modestly increase our sample size, we have focused on developing the novel methodology we use here. All tracer particles within 500 $h^{-1}$ kpc of the subhalo's $z=0$ location are then selected to be followed through cosmic history. This selection ensures that essentially any particle that has ever interacted with the galaxy's disk is included. Each tracer particle's unique ID is then found (or not) in every subbox snapshot, where the position, velocity, gravitational potential, particle type, internal energy (for gas), and age (for collisionless particles) of the tracer's instantaneous host are recorded. The particles' positions and velocities are transformed into a cylindrical coordinate system centered on a time-smoothed estimate of the galaxy's baryonic center of mass and velocity, with a z-axis oriented along the baryonic angular momentum. Details of how the reference frame of the galaxy is traced back to high redshift and smoothed are discussed in Appendix~\ref{app:smoothing}.

\begin{figure}[h]
    \centering
    \includegraphics[width=0.8\textwidth]{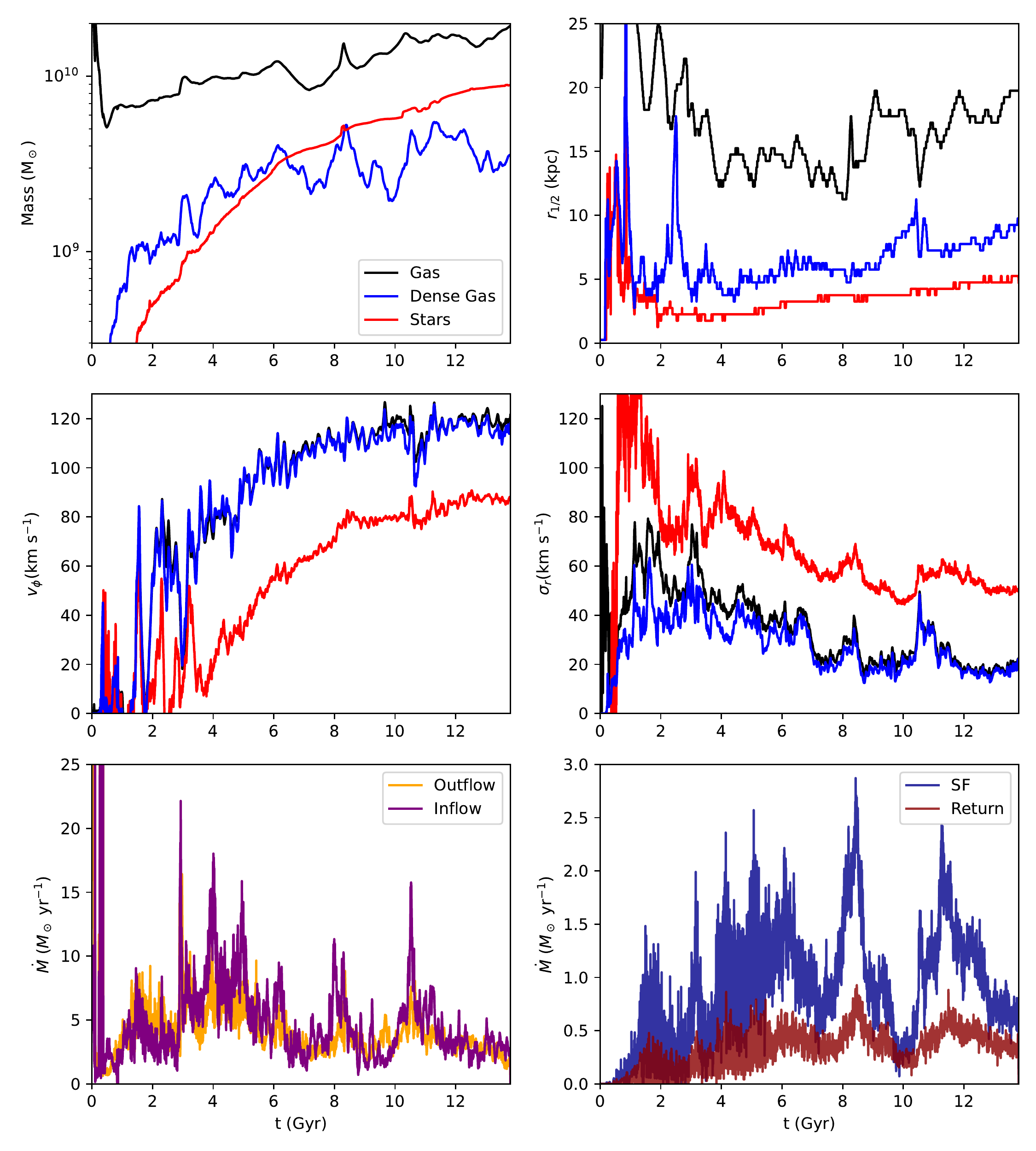}
    \caption{Properties of the galaxy over cosmic time. The first 4 panels show these quantities for tracer particles attached to the gas phase, stars, and dense gas, defined as gas with densities exceeding 0.1 cm$^{-3}$, approximately the threshold for star formation. The mass in each component, the half-mass radius $r_\mathrm{1/2}$, and the mass-weighted average $v_\phi$ and $\sigma_r$ within 20 kpc of the galactic center are shown. The profiles of these and the other velocity components are shown in the following figure. We also show the inflow, outflow, star formation rate, and return rate within 20 kpc in the final two panels.}
    \label{fig:galaxyProperties}
\end{figure}

\begin{figure}[h]
    \centering
    \includegraphics[width=0.8\textwidth]{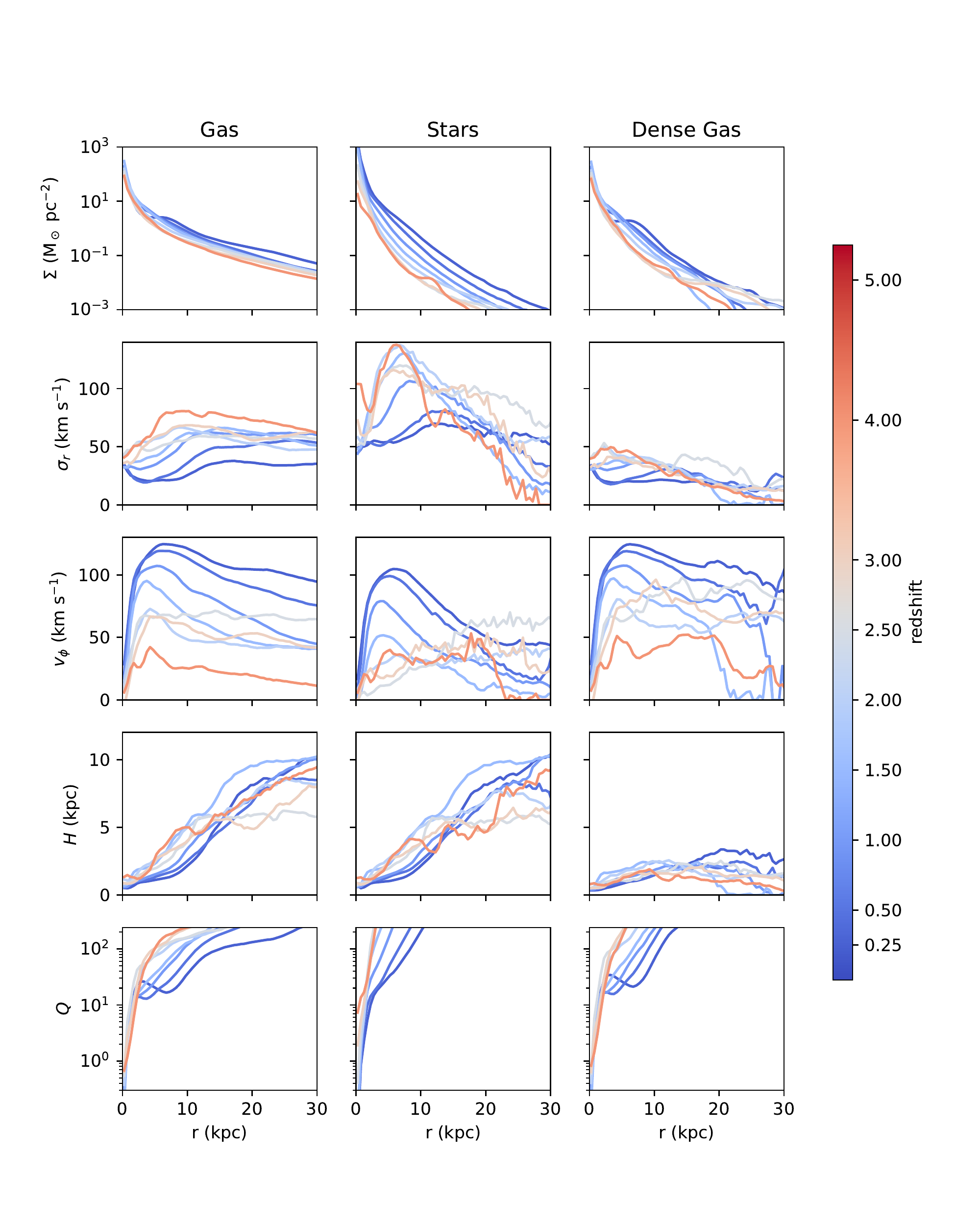}
    \caption{Radial profiles. For each component (columns), we show the average radial profile in bins of redshift (color) for different quantities (rows). The mass per unit area, velocity dispersion, mean velocity, scale height, and Toomre $Q$ parameter for each component are averaged over all snapshots within the specified redshift bin -- see the text for details of how each quantity is computed. Note that these profiles extend to radii substantially larger than what one might traditionally define as the extent of the galaxy, which for this relatively low-mass galaxy would be $\sim 10 $ kpc, i.e. $\sim 2$ times $r_\mathrm{1/2}$ in stars.}
    \label{fig:radialProperties}
\end{figure}

\begin{figure}[h]
    \centering
    \includegraphics[width=0.95\textwidth]{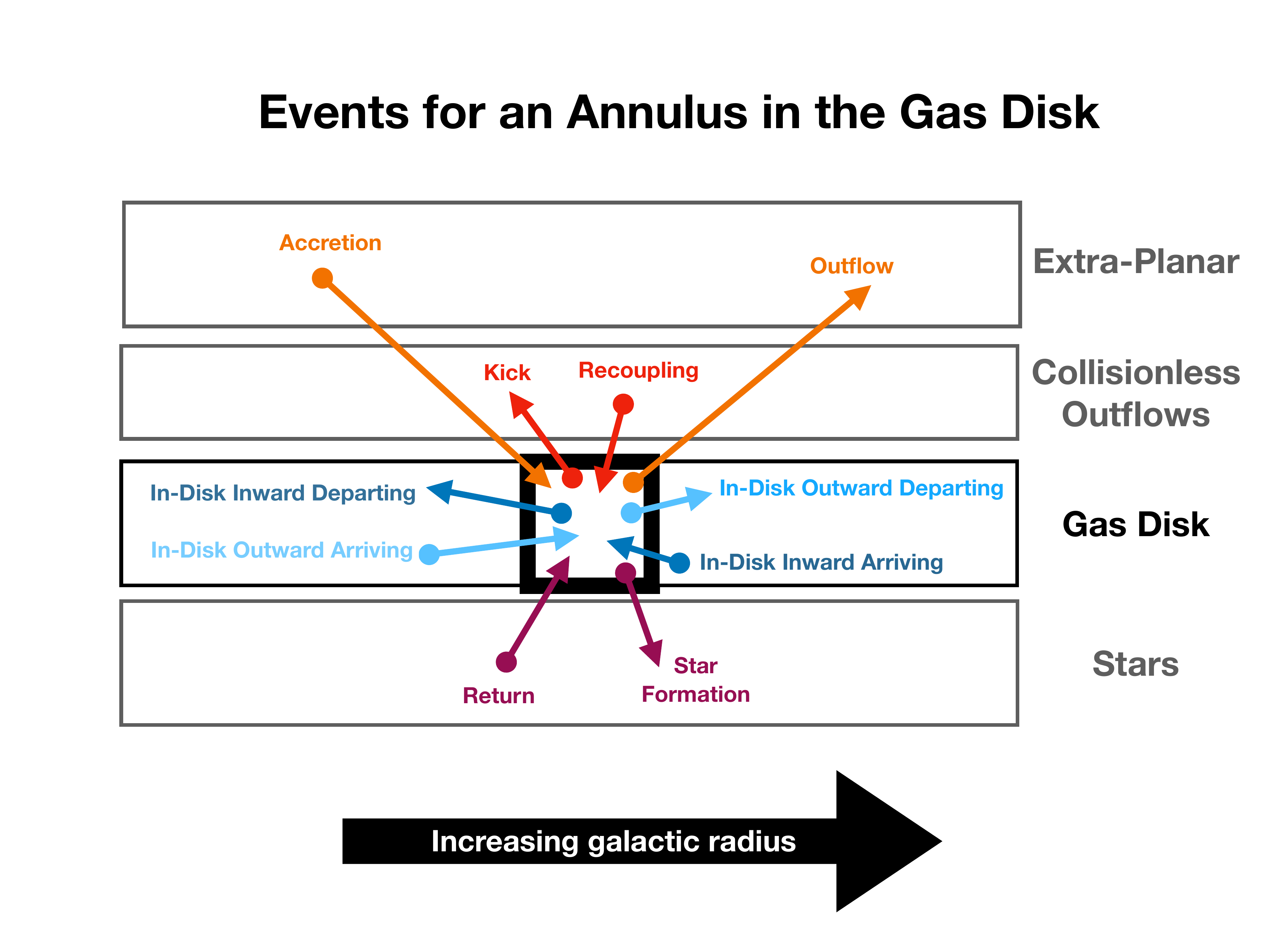}
    \caption{Event typology. The black square represents an annulus within the gas disk, and the arrows show different events that a given tracer particle may undergo between one timestep and the next. The base of the arrow shows the particle's origin in the previous snapshot, and the head shows where the tracer particle ends up in the later snapshot. For Kick, Accretion, Recoupling, Outflow, Return, and Star Formation events, the exact location of the particle outside the annulus in question is not relevant for classifying the event as such, e.g. the star particle formed in a Star Formation event may in principle be outside of the disk or in the same annulus in the later snapshot. All that matters is that the particle is changing whether or not it is associated with {\it gas in this annulus}. With the exception of the ``In-Disk'' events, the direction of the arrows is not meant to be literal.}
    \label{fig:eventsCartoon}
\end{figure}

With the galaxy selected and a frame of reference for the galaxy's orientation chosen, we now turn to briefly examining the basic properties of this galaxy. First, simple histogram-based images of the tracer particles associated with the gas and stars (blue and red respectively) at redshifts 3, 2, 1, and 0 are shown in Figure \ref{fig:postage}. Unsurprisingly, the galaxy only roughly resembles a disk at high redshift, before settling into a thin disk with prominent spiral structure in the gas. 
Basic properties of the galaxy are shown in Figure \ref{fig:galaxyProperties}, and several radius-dependent quantities are shown in Figure \ref{fig:radialProperties}. Radius-dependent quantities are computed by dividing the galaxy into 0.5 physical kpc annuli at each subbox snapshot and, for the star-forming gas, the stars, and all gas, computing column densities, velocity dispersions, and scale heights. These reference quantities augment the basic galaxy properties and will play a role in the subsequent analysis. First, we estimate the vertical density profile of each component in the annulus by selecting all tracer particles of that type with a cylindrical radius in the appropriate 0.5-kpc range, and with $|z|$ less than 10 kpc or the cylindrical radius, whichever is greater. This generous selection is designed to prevent any edge effects near the boundary of the selection. A kernel density estimate of the density is then constructed with a bandwidth estimated by Scott's rule \citep{scott_multivariate_2015}. The maximum density is taken to be the location of the midplane, and the first point above and below the midplane where the density falls to $\mbox{sech}^2(1.0)\approx 0.42$ of the maximum density are recorded. The scale height $H$ is estimated as the distance between these two points, divided by two. Mean velocities and their standard deviations are then computed for all particles within $\pm H$ of the midplane location in each of the three cylindrical coordinates. In practice these velocities are not sensitive to this selection, since in general most of the mass will fall into this region, and any selection that selects most of the mass will yield similar results. Finally, column densities are taken to be simply the number of particles within the appropriate annulus, and within a distance $|z|$ of 3 kpc or $r/3$, whichever is greater. Again as long as this selects most of the mass in the annulus, there is little sensitivity to this choice. The corresponding quantities shown in Figure \ref{fig:galaxyProperties} are mass-weighted averages of each component (dense gas, i.e. gas with densities above 0.1 cm$^{-3}$, all gas, and stars) over the central 10 kpc of the disk at each snapshot.

Each event is taken to occur at a particular snapshot, for a particular Lagrangian tracer particle, and is associated with an annulus in the gas disk. We identify the following types of events (see Figure \ref{fig:eventsCartoon})
\begin{enumerate}
    \item {\bf Star Formation}, where the tracer particle's parent's type changes from gas to a collisionless star particle).
    \item {\bf Return}, where a tracer particle moves from a star particle to the gas phase to model mass loss over the course of stellar evolution.
    \item {\bf Outflow}, where a particle tracing gas leaves the 3D volume that defines the disk.
    \item {\bf Accretion}, where a particle tracing gas enters the 3D volume that defines the disk.
    \item {\bf In-disk \{Inward/Outward\} \{Arriving/Departing\}}, where a particle tracing gas arrives in or departs from the gas disk annulus in question, respectively arriving from or departing to another annulus in the gas disk.
    \item {\bf Kick}, where a particle exits the gas phase to become a collisionless outflow particle used in the TNG model to capture the production of galactic winds due to stellar feedback.
    \item {\bf Recoupling}, when a collisionless outflow particle recouples to the gas disk, imparting energy, momentum, and mass within the volume of the disk.\footnote{In general outflow particles will recouple to the gas outside the disk, but this is not guaranteed, especially given the somewhat expansive definition of disk volume that we employ here. These recoupling events therefore play some direct role in the energy budget of the disk.}
    
\end{enumerate}
The events take place ``between'' two snapshots, in that each event type is defined by a Lagrangian tracer's parent particle changing state in some way between one snapshot and the next. We assign each event to its location within the disk annulus, which means that for events departing that annulus, we use the particle's position in the earlier snapshot, while for events arriving in the annulus, we use the particle's position at the later snapshot. We then assign the time of the event to the earlier snapshot. These choices are made following the principle of detailed balance, in the sense that with these choices, the number of gas particles in every particular annulus is exactly equal to the cumulative sum of the number of ``arriving'' events (Return, Accretion, Recoupling, In-Disk Inward Arriving, In-Disk Outward Arriving) minus the cumulative sum of all ``departing'' events (Star Formation, Outflows, Kicks, In-Disk Inward Departing, In-Disk Outward Departing), plus the initial number of gas particles within that annulus. This will be convenient later when we try to understand the energy budget of individual annuli.


\section{Results} \label{sec:res}

We now employ four separate methods to understand the energy budget of each annulus in the disk. First, we stack like events in bins of redshift to see how the energy of the particles undergoing these events change before and after the event. Then, we fit an explicit model for the total energy in an annulus as a function of time, taking advantage of our knowledge of the exact number of events of each type in each timestep. These methods provide estimates of the energy {\it budget}, but do not necessarily establish how efficiently the available energy is converted from its original form to turbulence in the simulation. To uncover potential causal relationships, we examine cross-correlation functions between the  energy in the gas velocity dispersion and time series of other quantities, looking for asymmetries around the point of zero time lag, which strongly suggest causal relationships. Finally, we fit a forward model for the turbulent energy only and check how it compares with the model for the total kinetic energy.

\subsection{Total Energy Budget - Stacking}

\begin{figure}[h]
    \centering
    \includegraphics[width=0.8\textwidth]{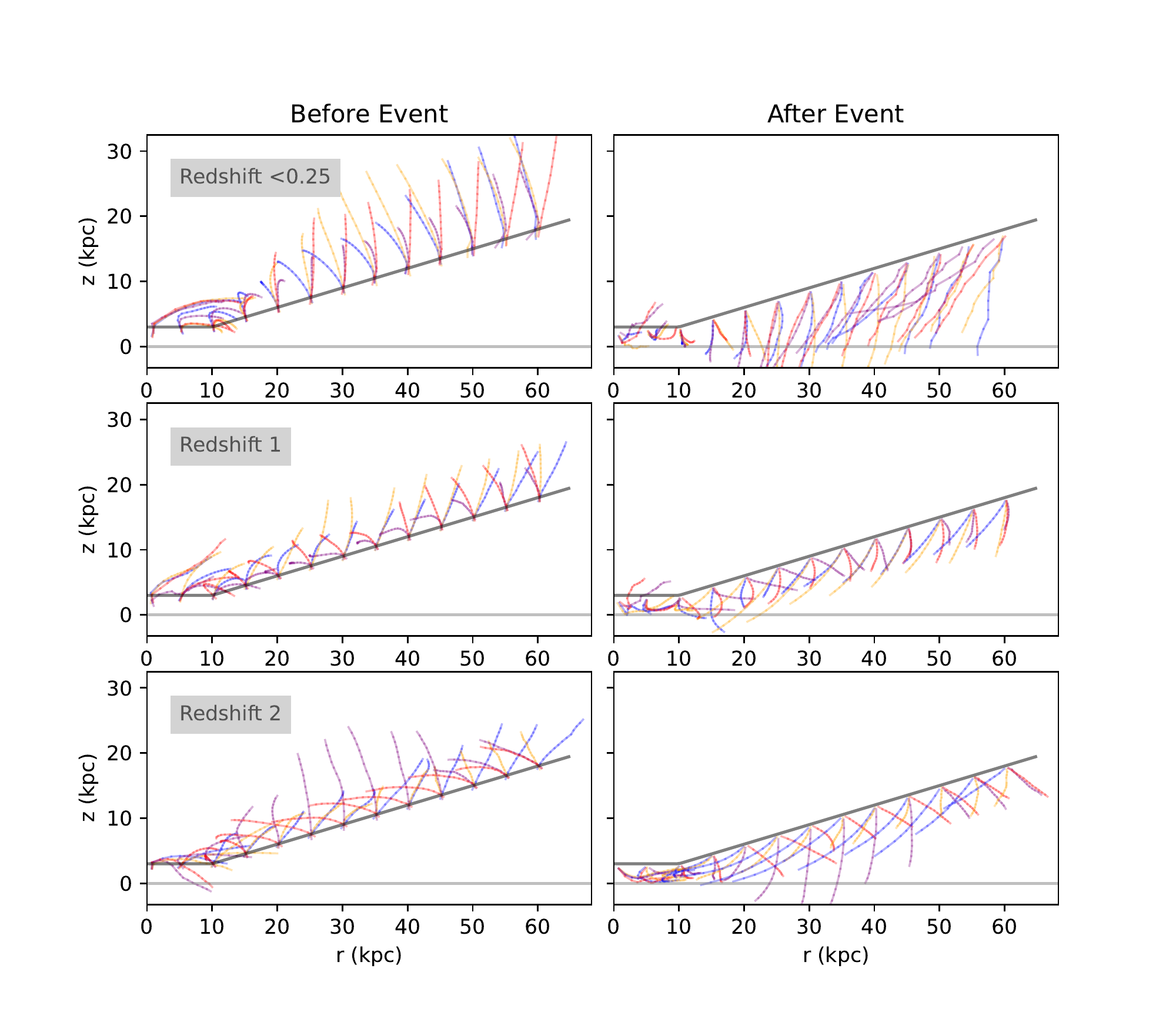}
    \caption{Average trajectories of accreting particles in $r$ vs. $z$. The left and right columns show the trajectories prior to and after, respectively, the accretion event. The different colors show the 4 $\phi$ bins -- around the cylinder they are colored blue, orange, purple, and red. The black lines show the disk volume that is used to define whether a particle is inside or outside of the disk for the purpose of identifying the events illustrated in Figure \ref{fig:eventsCartoon}.}
    \label{fig:rz}
\end{figure}

\begin{figure}[h]
    \centering
    \includegraphics[width=0.8\textwidth]{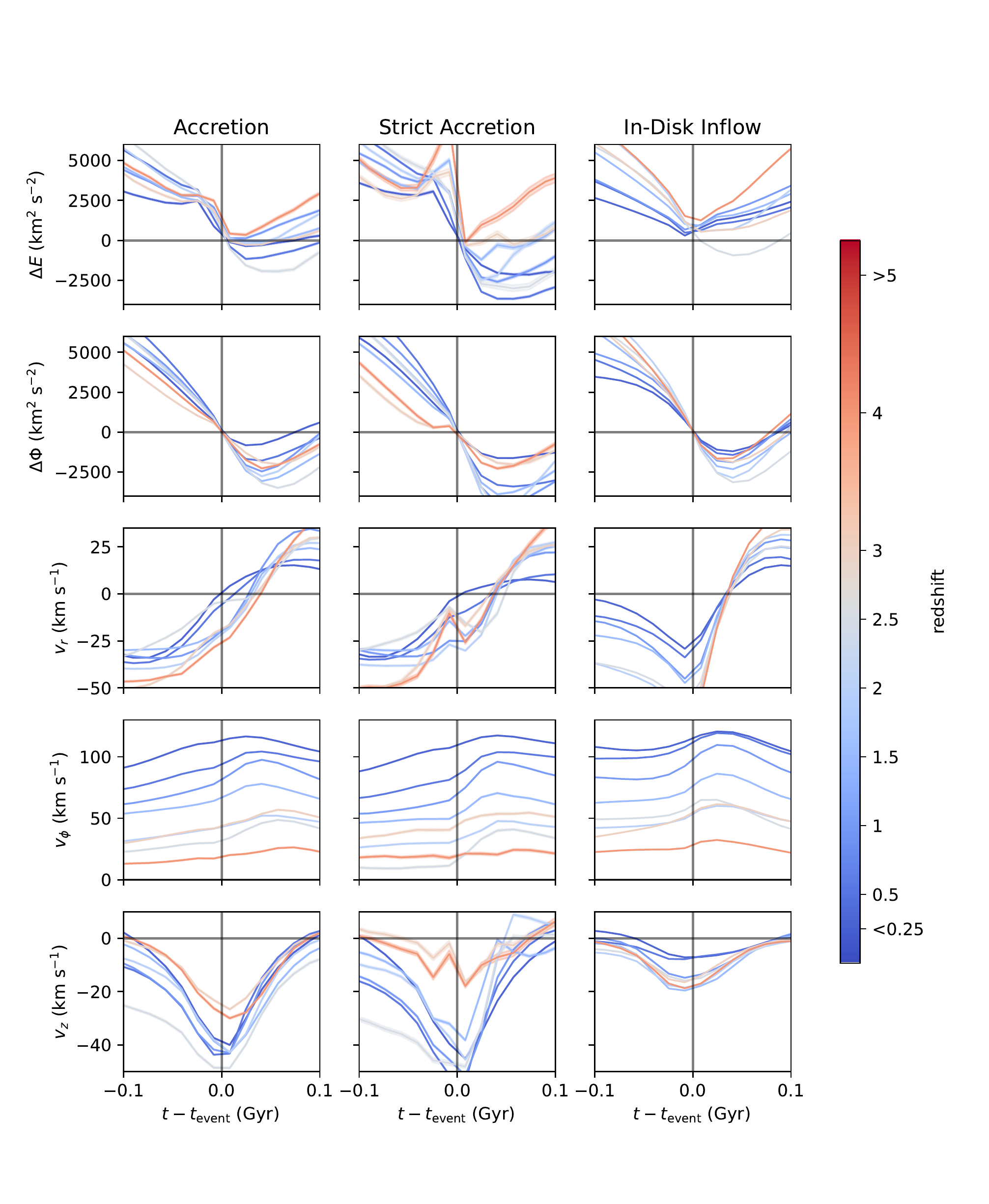}
    \caption{Average properties of accreting particles, accreting particles with additional cuts, and particles inflowing within the disk. The average of each property is taken at a sequence of times before, during, and after each event, limited to events that occur within 10 kpc of the galactic center. This technique may be considered a form of ``stacking,'' where instead of images or spectra, the quantity being stacked is the property of the tracer particle as it experiences the given event. Standard errors on the mean are shown as shaded regions, but are usually too small to see due to the large number of events. The colors indicate the redshift bin of the events. In order, the quantities plotted are $\Delta E$, the change in Kinetic plus Potential energy of the particle relative to the time of the event, $\Delta \Phi$, the change in potential energy of the particle relative to the time of the event, and the three velocity components $v_r$, $v_\phi$, and $v_z$ in the cylindrical coordinate system of the galaxy. Note that $v_z$ is defined such that negative velocities 40 Myr before the event mean that the particle's z-velocity is towards $z=0$ at that time.}
    \label{fig:tracks}
\end{figure}

In bins of radius, $\phi$ (i.e. the azimuthal angle about the z-axis of the galaxy), redshift, and time before and after each event, we record the average and standard error on the mean of the physical properties of the tracer particles as enumerated in the previous section (i.e. their positions, velocities, energies, and densities in the frame of the galaxy). Each event type is recorded separately. We can therefore examine quantities like the average change in radius before and after a particle accretes onto the galaxy. We use 4 bins in $\phi$, 130 bins in radius evenly spaced from 0 to 65 kpc, 50 evenly-spaced time bins around the event, from 400 Myr prior to 400 Myr after the event, and 8 bins of redshift at $z<0.25$, and around redshifts 0.5, 1, 1.5, 2, 2.5, 3, and 4. 

The average $r$ and $z$ coordinates of accreting particles during the 100 Myr prior to their time of accretion is shown in Figure \ref{fig:rz} in several different redshift bins. The 4 colors show the averages for different $\phi$-bins. The left columns show the 10 Myr prior to the accretion event, and the right columns show the 10 Myr after. The black line shows the boundary of the disk volume, namely $|z|$ values less than 3 kpc or $r/3$, whichever is greater, which is used to define inflow and outflow events. To avoid averaging to $\sim$ zero in quantities like $z$ and $v_z$ which {\it a priori} possess two-fold symmetry about the plane of the galaxy, each particle's contribution to the moments of a trajectory may be multiplied by $-1$ as follows. For $z$, we use the convention that for accretion events, at 40 Myr prior to the event, $z>0$, and $v_z < 0$. Earlier or later parts of the trajectory may change signs of course if the particle crosses the midplane during this time, or changes $z-$directions respectively. This means that rather than simply assigning $v_z<0$ to mean a velocity towards the disk, $v_z<0$ means the particle's $z$-velocity is the same as it was at the reference time. This definition, therefore, does not force $v_z$ to change at the midplane if the particle has not changed its actual direction of motion. 

The particle tracks show several qualitatively different behaviors. In the inner galaxy, particles tend to be arriving from further out in the disk, and in fact coming from the disk itself, with many average trajectories originating from the volume defined by the disk. In the outskirts of the disk, the trajectories are much more vertically-oriented, becoming progressively more so at lower redshifts. In all cases, there is a substantial azimuthal dependence, which changes from redshift bin to bin, e.g. the red lines preferentially lean towards lower r at $z\sim1$, but higher $r$ at $z\sim 0$. The fact that the purple trajectories at $z \sim 2$ all seem to line up around $\sim 30$ kpc suggests the tracks may be dominated by small numbers of substantial inflow events, which appears to be qualitatively true in movies of this galaxy.

Beyond just $r$ and $z$ trajectories, we can of course examine the average trajectories of other quantities. In Figure~\ref{fig:tracks}, we show the change in energy (kinetic plus gravitational potential), gravitational potential energy, and the values of the three velocity components in cylindrical coordinates, again keeping in mind the convention that $v_z\le 0$ just prior to the time of the event.  In addition to showing (in the left column of Figure \ref{fig:tracks}) these trajectories for accretion events, namely all events where a gas particle enters the volume defined by the disk between two adjacent snapshots, we also show these trajectories for a stricter definition of accretion in the middle column of the same figure. In particular, this definition requires in addition that the particle's density increase by a factor of 3 at some point between the event and the end of the 100 Myrs following the event, and similarly that the $z-$velocity must decrease in magnitude by at least a factor of 3. The intention is to exclude particles that pass through this region without interacting with anything resembling a disk, as may be the case for gas bound to satellite galaxies or locations with very low midplane densities. This stricter definition of accretion is used here just as a point of comparison -- everywhere else we use the simple geometric definition to preserve the ``detailed balance'' of each gas annulus, namely that the number of particles in each annulus is exactly equal to the sum of all events adding a particle to the annulus minus the sum of all events removing a particle. The strict and simple definitions of accretion yield similar trajectories in these parameters, particularly at lower redshifts, though with about a factor of two greater magnitudes in energy.

From the perspective of the energy budget, it is clear that accreting particles experience an inflection point in $\Delta E(t)$ around 10 Myr prior to the accretion event, which suggests that a few thousand km$^2$ s$^{-2}$, which for this galaxy is comparable to (about half of) $(1/2) v_\mathrm{circ}^2$, of specific energy is shared with the ambient gas. For strict accretion events, the energy budget is several times larger per particle. \citet{klessen_accretiondriven_2010} suggest that the energy available to drive turbulence in the case of gas accreting onto a galaxy should be $\sim \epsilon v_\mathrm{circ}^2$, where $\epsilon$ is an efficiency factor comparable to the density contrast between the accreting gas and the object onto which it is accreting. The density of the particles before and after accretion events (see the dashed lines in Figure \ref{fig:efficiency}) shows that on average the density contrast is only about a factor of a few (and of course more for particles whose trajectories were selected in part based on their density contrast). These tracks of the particles' energy alone do not necessarily give us much information about $\epsilon$, but the energy budget does agree with the simple expectation of $(1/2)v_\mathrm{circ}^2$.

Since we can look at not just the energy, but the change in the gravitational potential, and the change in each velocity component, we can see that the change in energy on $\sim 100$ Myr timescales largely follows the change in the potential energy. Particularly at $z<4$ and when considering just the geometrically-defined accretion events, the inflection on $\lesssim 10$ Myr timescales around the time of the event is not present in the potential energy. It is also not present in any single velocity component, meaning that the energy is at least in part ending up elsewhere, i.e. some combination of thermal energy, radiative losses, or turbulence.

\subsection{Total Energy Budget - Explicit Model}

\begin{figure}[h]
    \centering
    \includegraphics[width=1.0\textwidth]{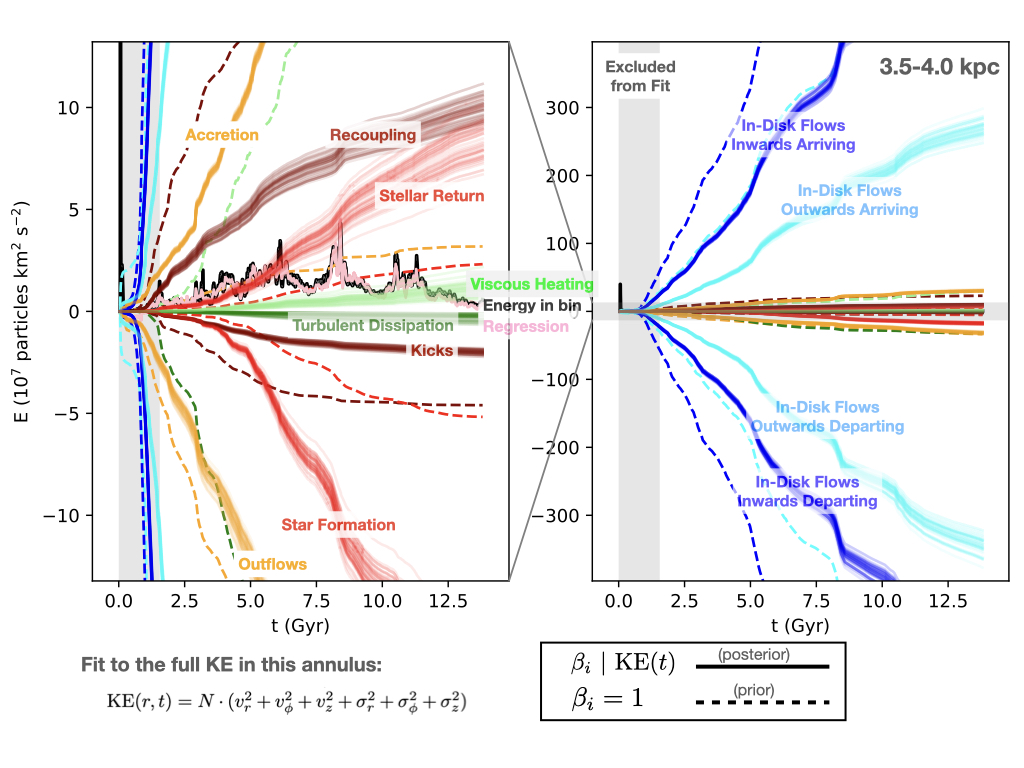}
    \caption{An illustration of the fitting procedure in one particular annulus of the disk. The black line shows the actual kinetic energy in this annulus estimated as the sum of the energies associated with the mean and dispersion of the $r$, $\phi$, and $z$ cylindrical components of the velocity, and with mass measured in units of number of tracer particles, each of which is associated with $8.5 \times 10^4 M_\odot$. The series of solid lines of various colors show posterior samples for that quantity. For instance, the red lines show the range of energy contributions from Star Formation (negative) and Return (positive) events as inferred by the fit. The corresponding dashed lines show the contribution from these events if their corresponding $\beta_i$ were set to 1, i.e. the prior average. The pink lines show draws from the posterior for the sum of all terms. The left panel is a zoomed-in version of the right panel, since the terms have such large variation in amplitude.}
    \label{fig:singlecolumn}
\end{figure}

\begin{figure}[h]
    \centering
    \includegraphics[width=0.95\textwidth]{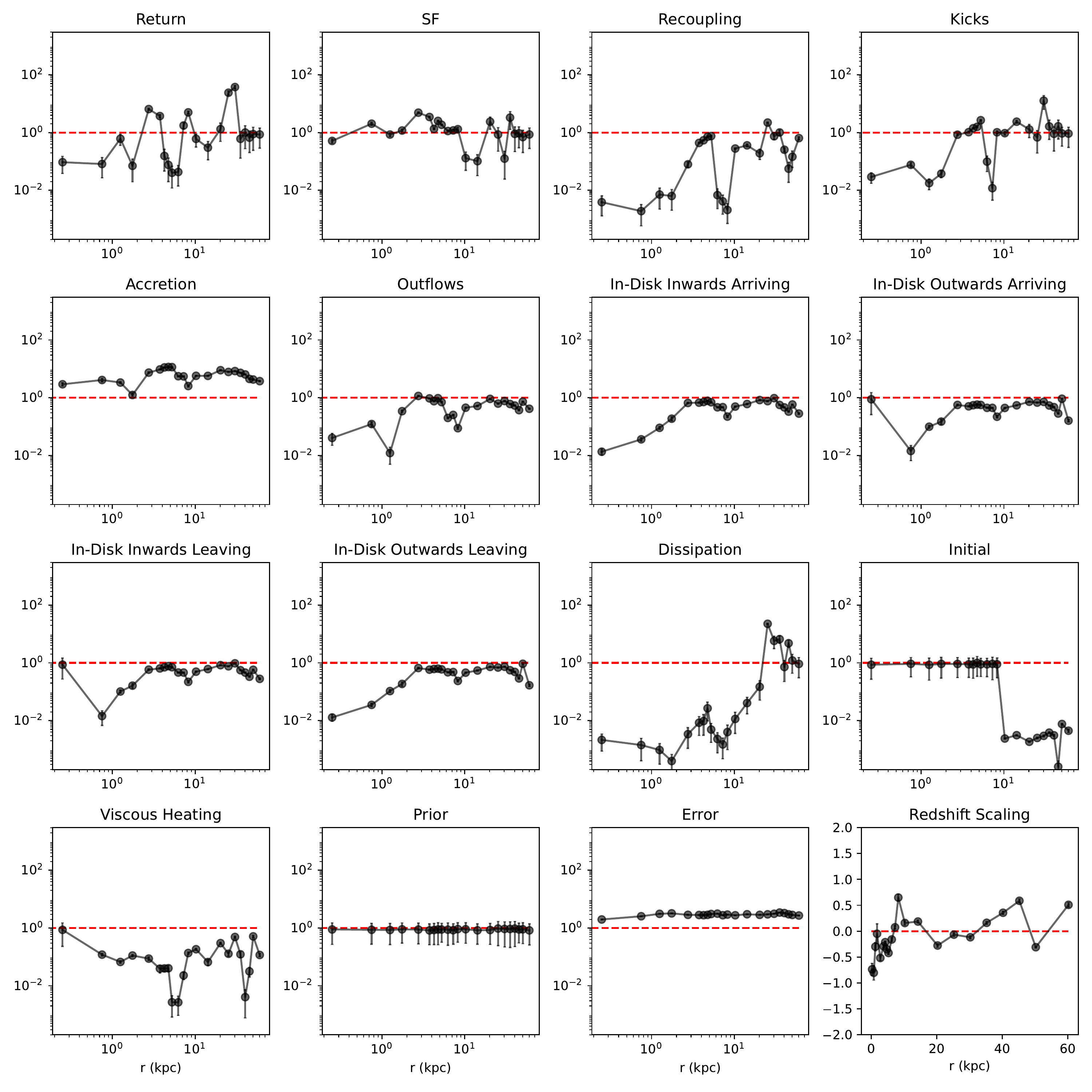}
    \caption{The coefficients and their error bars, i.e. the $\beta_i$, fit to the data following equation \ref{eq:KE}. Each panel shows a different $\beta_i$ as a function of radius (note that the fit at each radius is independent from the other radii). The red dashed lines show the center of the prior. The first 10 panels show the $\beta_i$'s corresponding to each of the event types shown in Figure \ref{fig:eventsCartoon}. The $\beta$ parameters for the Dissipation, Viscous Heating, and Initial Condition terms are shown next. The Prior panel represents how the full prior (not just its mean) appears in this format. The Error panel shows the $\beta_i$ term controlling the magnitude of the Error on the energy in this bin (see Equation \ref{eq:errorterm}). Since the fits at each radius are computed totally separately, the smooth behavior as a function of radius for many of these parameters suggests that enough information exists in the data at each radius to produce a reasonable estimate for these quantities.}
    \label{fig:efac}
\end{figure}

In the previous subsection, we examined the average properties of tracer particles as they underwent accretion events. Their energies as a function of time suggest that they are able to share of order $v^2_\mathrm{circ}$ per particle of their energy with the disk when they join it. However, it is not clear what form this energy takes, nor whether the energy budget of individual tracer particles calculated in this way is sufficient to understand the energy budget of the disk itself. To address these questions, and see whether the high accretion efficiencies suggested by the results of the previous subsection hold up, we can take advantage of our careful partitioning of every tracer particle's entrance or exit into individual annuli in the disk. This allows us to build an explicit forward model for the energy balance in each annulus separately, allowing us to see radial trends, and account for terms in the energy budget that are not linearly proportional to any particular mass flow, most notably (turbulent) dissipation.

In each of several annuli, we estimate the total kinetic energy as a function of time (in units of tracer particles km$^2$ s$^{-2}$) as $\mathrm{KE}(r,t) = N \cdot (v_r^2 + v_\phi^2 + v_z^2 + \sigma_r^2 + \sigma_\phi^2 + \sigma_z^2)$, i.e. the sum of the square of the average velocity in each component and the variance in each velocity component. We make the simple assumption that the kinetic energy in the annulus is the sum of many separable components, mostly associated with the mass fluxes from individual event types. In particular, we assume that these terms are close to linear, so that the energy contribution from, say, ``recoupling'' events, is equal to the rate of such events times a particular specific energy $(1/2) v^2$, that changes in a pre-set, i.e. non-tunable, way over the course of the galaxy's life. Under this assumption, we take the kinetic energy to be the following sum over event types $i=1...S$,
\begin{equation}
   \mathrm{KE}(r,t) = \beta_{S+3} N_i(0) (10\ \mathrm{km}\ \mathrm{s}^{-1})^2 +  \left( \sum_{i}^S \beta_i \int_0^t \dot{N}_{i}(t) (1/2) v_{i}^2 dt + \int_0^t (\beta_{S+1}\dot{E}_\mathrm{visc} - \beta_{S+2}\dot{E}_\mathrm{diss}) dt\right) (1+z)^{\alpha_{z}} + \epsilon_{\mathrm{const}} .
   \label{eq:KE}
\end{equation}
Here $\dot{N}_i(t)$ is the change in the number of tracer particles in the annulus due to event type $i$, and $v_i$ is the velocity associated with that event type, discussed below. 
The model parameters, the $\beta_i$ for $i=1, ..., S+3$ 
have\footnote{In addition to the $S+3$ $\beta_i$ parameters explicitly shown in equation \ref{eq:KE}, there is a parameter $\beta_{S+4}$ that controls the variance of the error term $\epsilon_\mathrm{const}$, defined in Equation \ref{eq:errorterm}. } 
gamma-distributed priors with mean $1$ and variance $0.4$. This prior requires that $\beta_i>0$ and allows the parameters to be larger than $1$, but ensures that most of the probability lies below $1$, reflecting the fact that the terms multiplying the $\beta_i$ are generally rough estimates of the energy budget associated with a physical process, meaning that most of the time they should be a number of order unity, but usually less. The $S=10$ types of events are the same as shown in Fig. \ref{fig:eventsCartoon}, so the $\dot{N}_i$, which can be positive or negative, account for all of the flows into or out of the annulus\footnote{In principle tracer particles could move into and out of an annulus between two snapshots, and thereby affect an annulus without being recorded as an event associated with that annulus. This is rare in the simulation, at least in the subbox with its closely-spaced snapshots.}. 

Each event type has an associated velocity $v_i$ that we estimate with physical considerations. Unless otherwise noted, $v_i = v_i(t) = \max_r(v_\mathrm{circ}(r,t))$, i.e. the maximum circular velocity within the 70 kpc covered by the cylindrical annuli. The exceptions are Return, Star Formation, Kicks, and Accretion. For accretion we adopt $v_i = \sqrt{0.1}\max_r(v_\mathrm{circ}(r,t))$ since we expect {\it a priori} that there should be an efficiency factor of that order. Kicks and Star Formation act on star-forming gas only, so we use $v_i^2 = v_{r,\mathrm{SF}}^2 + v_{\phi,\mathrm{SF}}^2 +v_{z,\mathrm{SF}}^2 +\sigma_{r,\mathrm{SF}}^2 +\sigma_{\phi,\mathrm{SF}}^2 +\sigma_{z,\mathrm{SF}}^2$, where the subscript SF indicates that the measurement is made in the star-forming gas only. For Return events, we average the specific kinetic energy associated with star-forming gas with the specific kinetic energy associated with the stars, since Return events are a mixture of almost-immediately returned gas (e.g. from supernovae) and gas returned much later in the stellar population's life (e.g. AGB winds). To allow the model some additional flexibility, the parameter controlling the $z-$scaling, $\alpha_z$ has a normal prior centered on zero with standard deviation $0.5$. Physically this parameter may be a proxy for gradual changes in efficiency in the parameters or non-linearities in the effect of each event, though there is no reason to expect that all such changes or non-linearities should be the same for all of the terms. As a practical matter, this parameter just serves to ease some of the rigidity of the model assumptions.

Besides the flow-based terms, we expect that random motion may be generated by gravity, and may be dissipated by radiative cooling, giving rise to the terms $\dot{E}_\mathrm{visc}$ and $\dot{E}_\mathrm{diss}$ respectively. In the formalism of a thin viscous disk \citep{shakura_reprint_1973, shu_physics_1992, krumholz_dynamics_2010}, the viscous heating is
\begin{equation}
    \dot{E}_\mathrm{visc} = \frac{1}{2\pi r}\left(\frac{\partial\ln v_\mathrm{circ}}{\partial\ln r} -1\right) v_\mathrm{circ} \mathcal{T}
\end{equation}
where $\mathcal{T} = 2\pi r^2 \int T_{r\phi} dz$ is the vertically-integrated r-$\phi$ component of the viscous stress tensor $T$, and is related to the mass flow through the disk via the conservation of angular momentum as
\begin{equation}
    \frac{\partial \mathcal{T}}{\partial r} = \dot{M} v_\mathrm{circ} \left( 1 + \frac{\partial \ln v_\mathrm{circ}}{\partial \ln r} \right).
\end{equation}
We therefore estimate the fiducial $\dot{E}_\mathrm{visc}$ term as 
\begin{equation}
    \dot{E}_\mathrm{visc} = -\frac{1}{2\pi r} v_\mathrm{circ} \int_0^r \mathcal{I}_{\dot{N}<0} \dot{N}_\mathrm{in-disk} v_\mathrm{circ} dr'.
\end{equation}
Since $\partial \ln v_\mathrm{circ} / \partial \ln r$ should be close to $0$, and the energy budget at each radius is fit separately anyway, we neglect these factors in the fiducial estimate to avoid introducing a source of error from numerically differentiating a noisy quantity. We have also implicitly assumed the boundary condition $\mathcal{T}(r=0) =0$, which agrees for instance with the \citet{krumholz_dynamics_2010} analytic solution. Finally, we have included an indicator function $\mathcal{I}_{\dot{N}<0}$, which is 1 when $\dot{N}_\mathrm{in-disk}<0$, i.e. the mass flows inwards in the disk, and 0 otherwise. Since this computation is done at every snapshot and the integral is carried out over all interior radii, this eliminates negative contributions to $\dot{E}_\mathrm{visc}$ which are unphysical. Meanwhile the dissipation term $\dot{E}_\mathrm{diss}$, defined as the rate of energy loss not associated with any particle flux, is estimated as
\begin{equation}
    \dot{E}_\mathrm{diss} = N (\sigma_r^2 + \sigma_\phi^2 + \sigma_z^2) \frac{\sigma_z}{H}.
\end{equation}
This formula reflects the classic result that turbulent energy decays on the crossing timescale of the driving \citep{maclowKineticEnergyDecay1998, stoneDissipationCompressibleMagnetohydrodynamic1998} scale, taken to be of order the gas scale height. As discussed in the previous section, $H$ is measured for the gas based on the vertical density of tracer particles in the annulus, with $H$ defined as the distance from the maximum vertical density where the density drops below $\mathrm{sech}^2(1)\approx 0.42$ of its maximum value.

The error term, $\epsilon_\mathrm{const}$ is assumed to be independently Gaussian-distributed at every snapshot with zero mean and a variance set by the final $\beta$ parameter $\beta_{S+4}$. In particular 
\begin{equation}
\label{eq:errorterm}
    {\rm  Var}(\epsilon_\mathrm{const}) = ( N \cdot (\beta_{S+4}\ 10\ \mathrm{km}\ \mathrm{s}^{-1})^2 )^2,
\end{equation} 
i.e. $\beta_{S+4}$ scales the typical velocity dispersion of the error around a fiducial value of $10\ \mathrm{km}\ \mathrm{s}^{-1}$. This simple model for the error has the advantage of being easy to interpret, but is otherwise somewhat arbitrary. 

An offset between the model and the data can arise from many sources, but the most prominent is probably the simplifying assumptions of the model itself. For instance, it is assumed that all energy contributions from particle flows are linear in the flow rate, whereas one might reasonably expect differences in the specific energy associated with a high-flow regime vs. a low-flow regime. The redshift-dependence of the model is best thought of as a proxy for something else, since redshift or scale factor are not physical quantities of any direct importance on the scale of a galaxy, so the particular form we have chosen here, one which affects all terms equally, is a reasonable guess but could obviously be generalized. We defer improvements to this model to future work, where ideally the contribution to the error from $\epsilon_\mathrm{const}$ would be reduced and supplanted by estimates for specific sources of errors, e.g. the shot noise associated with measuring the velocity dispersion based on a set of tracer particles.

An example of how these fits work is shown in Figure~\ref{fig:singlecolumn}. The actual kinetic energy in the annulus is shown as the black line. The basis functions, i.e. the terms in Equation \ref{eq:KE} when their corresponding $\beta_i=1$, are shown as dashed lines, while the posterior fits scaling these basis functions are shown as many series of solid lines. The posterior samples are obtained from emcee \citep{foreman-mackey_emcee_2013, foreman-mackey_emcee_2019}, with 100 walkers in the ensemble. For nearly all radii, the fits are well-converged. The sum of all terms in the fit is shown in pink, and matches the data (black) quite well. The highest-redshift points are excluded from the fit, since the energy budget there is set by processes unrelated to the physics of the disk, i.e. the expansion of the universe.  


The fitted coefficients across many different radii are summarized in Figure \ref{fig:efac}. The general continuity of most of the coefficients within the error bars is a good sign. We expect that the physics governing the specific energy associated with, say, tracer particles arriving in a particular annulus from elsewhere in the disk, should vary only slowly with radius. The fact that the coefficients usually behave this way suggests that the model is a reasonable approximation to what is actually happening in the simulation. The coefficients that have more irregular behavior with radius, particularly those corresponding to Return, Star Formation, Kicks, and Viscous Heating have several things in common that may contribute to their irregularity. First, the time series by which these coefficients are multiplied includes $v_i$'s that are derived from the data, so the $v_i$'s may vary on short timescales and from bin to bin owing to noise. Moreover these time series {\it a priori} are expected to contribute only modestly to the total energy budget (see Figure \ref{fig:singlecolumn}). One way to ameliorate this problem would be to fit all of the radii jointly in a hierarchical model, but since we are more interested in the quantities that behave regularly in radius, particularly the contribution from direct accretion, we do not attempt this here.

Interestingly, the fitted values for the $\beta_i$ corresponding to the dissipation term are all $\ll 1$ within 10 kpc, which we expect to have a substantial impact on the turbulent energy budget. In simplified models like that of \cite{krumholz_dynamics_2010} or the model we present later in section \ref{sec:disc}, essentially the only way turbulent energy is lost is via this term. If the dissipation is truly so small in the simulation, the turbulent energy is likely to be overestimated compared to real galaxies. 

Other substantial deviations from the prior include the $\beta_i$ for accretion, which is systematically higher than 1 by a factor of a few. Since the $v_i$ in this time series includes an efficiency factor of 10\%, it is not surprising that the posterior values imply a somewhat higher efficiency. The recoupling coefficient is systematically much lower than 1 throughout much of the disk, though this may be the result of the fact that recoupling, as designed, occurs far from star-forming regions, whereas the kinetic energy of each annulus to which we are fitting is measured within a single scale height of the midplane.

\subsection{Causality - Cross-correlations}
\begin{figure}[h]
    \centering
    \includegraphics[width=0.8\textwidth]{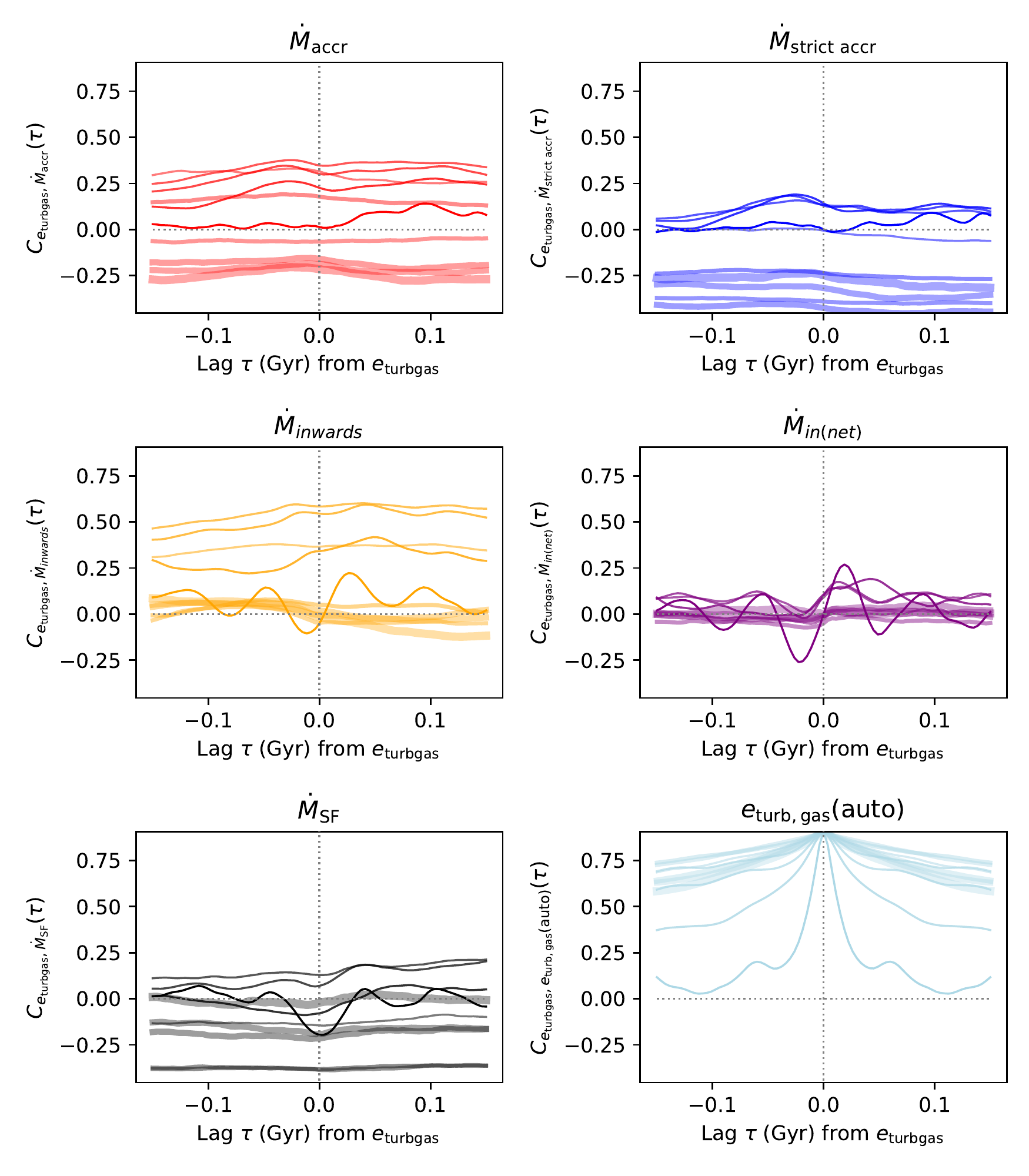}
    \caption{Cross-correlation functions as defined by Equation \ref{eq:crosscorr}. Lighter lines correspond to larger radii -- the radii shown are the annuli at 0.25 kpc, 0.75 kpc, 1.25 kpc, and 1.75 kpc, then every 2.5 kpc up to 20 kpc. In each case the specific turbulent energy of the gas is cross-correlated with another time series. From top left to bottom right, these are: the accretion rate, the accretion rate requiring kinematic- and density-based criteria in addition to the position of the particle simply entering the disk volume, the flow rate of tracer particles through the disk, the net flow rate of tracer particles through the disk, the star formation rate, and the specific turbulent energy itself. Negative values of the lag time ``lead up'' to features in $e_\mathrm{turb,gas}$, i.e. a positive cross-correlation at negative lag-time means that the quantity in question tends to be correlated with subsequent features in $e_\mathrm{turb,gas}$.}
    \label{fig:crosscorr1d}
\end{figure}

\begin{figure}[h]
    \centering
    \includegraphics[width=0.8\textwidth]{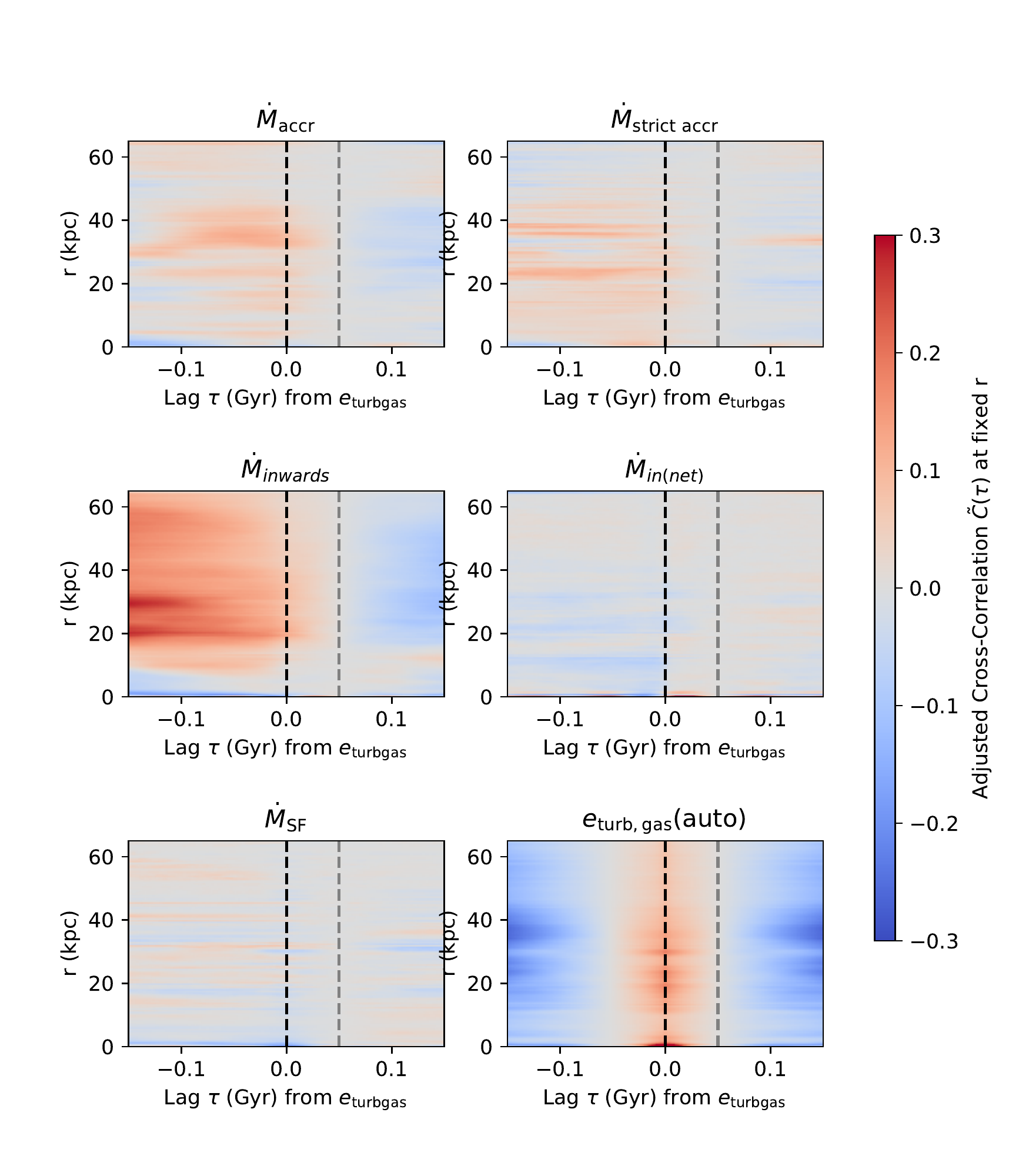}
    \caption{The same information as in Figure \ref{fig:crosscorr1d}, but the long-run correlations have been subtracted out. The adjusted cross-correlations $\tilde{C}(\tau)$ are set to zero along the gray dashed line at $\tau=50$ Myr. The strongly asymmetric values across $\tau=0$ suggest causal relationships between $e_\mathrm{turb,gas}$ and both $\dot{M}_\mathrm{accr}$ and $\dot{M}_\mathrm{inwards}$.}
    \label{fig:crosscorr2d}
\end{figure}

In the previous two subsections, we examined the energy available for each type of event identified with the tracer particles, both by stacking the trajectories of the particles in terms of their energy vs. time centered around each event, and by fitting a forward model to the energy contained in each annulus of the galaxy. In this and the following subsections we address a different question: does this energy translate into turbulence in the simulation? 

Before we fit another forward model for the turbulent energy (see the following subsection), we examine a non-parametric, model-independent statistic, namely the cross-correlation function. This is a generic method for understanding the relationship between two time series, and is defined by
\begin{equation}
    C_{f,g} (\tau) = \int_{t_\mathrm{min}}^{t_\mathrm{max}} \frac{(f(t) - \langle f \rangle) (g(t + \tau) - \langle g \rangle) }{\sigma_f \sigma_g} dt.
\end{equation}
The lag time $\tau$ between the two time series $f$ and $g$ may be positive or negative. The average and standard deviation of a time series $f$ are denoted $\langle f \rangle$ and $\sigma_f$, and the integral runs over the entire range of the time series.

In our case, the time series are sampled irregularly so each time series is first linearly interpolated and resampled at regular intervals equal to $(t_\mathrm{max} - t_\mathrm{min})/7200$, i.e. on average there are two points in the new grid for each snapshot. This preserves some of the very high cadence data at high redshift while still allowing the computation to proceed in a reasonable amount of time. We evaluate the lag time $\tau$ in 81 bins between $-300$ and 300 Myr. We also mask out times at high redshift ($z>4.5$), and outliers\footnote{The outliers in the timeseries will have an outsized effect on the cross-correlations, but are generally probing brief moments in certain annuli where the number of tracer particles drops to a low value.} in the various timeseries, namely $E_\mathrm{turb,SF}>7.5\times10^6\ \mathrm{particles}\ \mathrm{km}^2\ \mathrm{s}^{-2}$,  $e_\mathrm{turb,SF} > 5000\ \mathrm{km}^{2}\ \mathrm{s}^{-2}$, $\dot{N}_\mathrm{accr} > 20000\ \mathrm{particles}\ \mathrm{Gyr}^{-1}$, and $e_\mathrm{turb,gas} > 2\times 10^4\ \mathrm{km}^{2}\ \mathrm{s}^{-2}$. These quantities are defined intuitively: ``turb'' subscripts mean the kinetic energy from the random motions only, i.e. excluding the mean velocities. ``SF'' indicates that the quantity was calculated with the gas above the star formation threshold density, E and e refer to energy and specific energy respectively, and $\dot{N}_\mathrm{accr}$ is the number of accretion events arriving per unit time between any given pair of adjacent snapshots. If any of these conditions are met, the time is masked out for all time series, which in practice means that we estimate the cross-correlation between two time series derived from the simulation as
\begin{equation}
\label{eq:crosscorr}
    \hat{C}_{f,g}(\tau) = \left(\sum_i \mathcal{I}_{t_i} \right)^{-1} \sum_i \mathcal{I}_{t_i} \frac{(\hat{f}(t_i) - \langle \hat{f} \rangle) (\hat{g}(t_i + \tau) - \langle \hat{g} \rangle) }{\sigma_{\hat{f}} \sigma_{\hat{g}}}, 
\end{equation}
where the hats on $f$ and $g$ emphasize that these are the interpolated and resampled functions, the indicator function $\mathcal{I}$ is 1 at times that are not filtered out and 0 at times that are, and $i$ indexes the resampled times from 0 to 7199.

In Figure \ref{fig:crosscorr1d} we show the cross-correlation functions between the specific turbulent energy, i.e. $(1/2)(\sigma_r^2 + \sigma_\phi^2 + \sigma_z^2)$ associated with gas within 1 scale-height of the midplane $e_\mathrm{turb}$ and several other time series. Each line represents the cross-correlation between the two time series within a given radial annulus. Several phenomena are apparent: in the innermost annuli, represented by the thinnest lines, there is a characteristic oscillatory pattern on timescales of $\sim 30$ Myr. We speculate that this is the timescale associated with stellar feedback. Another common phenomenon is that radii slightly further out in the disk tend to show long-term trends. One possibility is that these long-term correlations reflect cosmological trends in the galaxy's growth, e.g. as the average rate of gas accretion slows down towards $z=0$, the star formation rate, mass flow rates, and the specific turbulent energy will all tend to decrease together. On the other hand, at large radii the dynamical time in the disk and CGM becomes substantially longer, so it may be that the largest radius bins in $\dot{M}_\mathrm{inwards}$ and $\dot{M}_\mathrm{accr}$ also reflect causal relationships between the quantities being compared.

If we remove these long-run trends by defining an adjusted cross-correlation function $\tilde{C} = \hat{C} - \hat{C}(50\ \mathrm{Myr})$, we will be able to see the correlation between these processes on shorter timescales more clearly. In Figure \ref{fig:crosscorr2d} we show $\tilde{C}(\tau)$ as a function of radius (each adjusted cross-correlation function is still calculated separately annulus-by-annulus). The asymmetry in these plots across a lag time of zero suggests a causal relationship between the two time series \citep{pierce_causality_1977, granger_investigating_1969, kubo_statisticalmechanical_1957}. The figure shows a very strong red-blue asymmetry for $\dot{M}_\mathrm{inwards}$ which changes sign near the center of the disk, and a weaker but still clear asymmetry for $\dot{M}_\mathrm{accr}$ and $\dot{M}_\mathrm{strict\,accr}$, while the diagrams for $\dot{M}_{\mathrm{in(net)}}$ and $\dot{M}_\mathrm{SF}$ are essentially uniform, and that for the autocorrelation of $e_{\rm turb,gas}$ is symmetric about $\tau = 0$. The clear interpretation is that on average, increases or decreases in $e_\mathrm{turb}$ tend to occur in the $\sim 100$ Myr after corresponding increases or decreases in $\dot{M}_\mathrm{accr}$ and $\dot{M}_\mathrm{inwards}$. We can have confidence that this is a real effect and not an artifact in large part because the signal disappears when looking at the cross-correlation of $\dot{M}_\mathrm{in(net)}$, i.e.~the mass growth at each radius as the result of mass flow through the disk. In other words, the turbulent specific energy depends on $\dot{M}_\mathrm{inwards}$, but \textit{not} on $\dot{M}_\mathrm{in(net)} = \partial \dot{M}_\mathrm{inwards} / \partial r$, which is exactly what we expect: the former is proportional to the energy injection rate as matter falls down the potential well, while the latter is proportional to part of the advection of kinetic energy through the disk \citep{krumholz_dynamics_2010}, and it is rare for the advection term to matter \citep{forbes_balance_2014}. Our interpretation is that, without reference to any explicit model, greater inward mass flux or direct accretion causes higher velocity dispersions. In contrast, there is relatively little effect from star formation.

\subsection{Causality - Forward Models for the Turbulent Energy}

\begin{figure}[h]
    \centering
    \includegraphics[width=.95\textwidth]{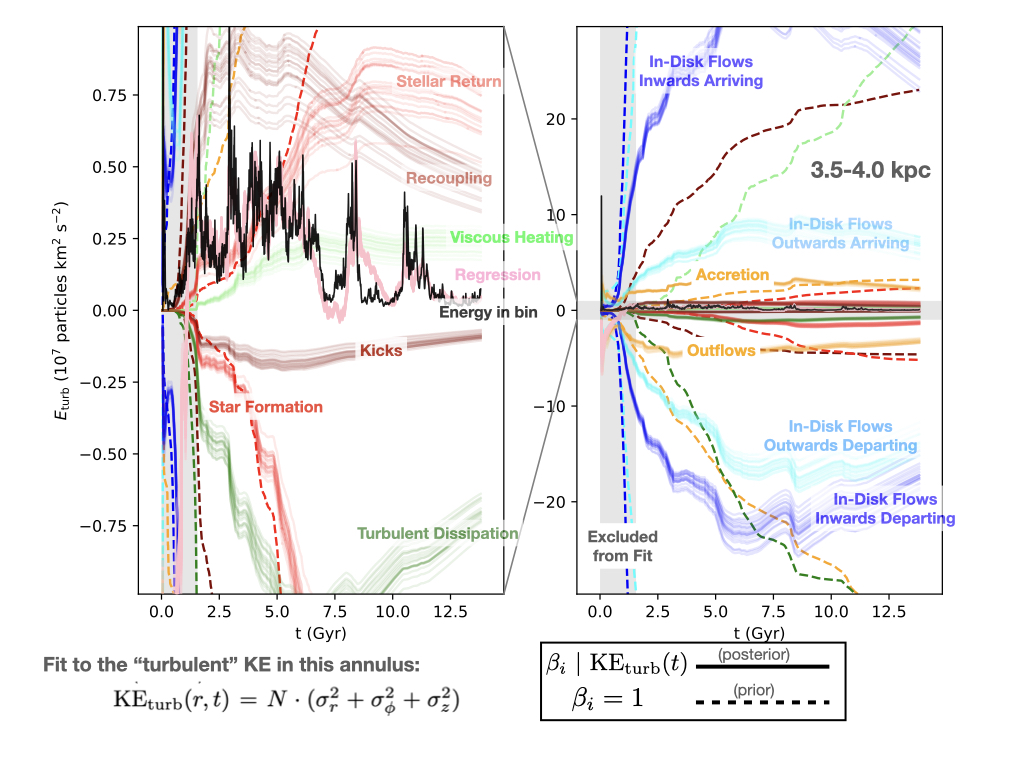}
    \caption{ Fits for the turbulent kinetic energy. Same as Figure \ref{fig:singlecolumn}, but fitting KE$_\mathrm{turb}(r,t)$ instead of $\mathrm{KE}(r,t)$. The key differences comparing this figure to Figure \ref{fig:singlecolumn} include that the energy scale is much lower, since only the random velocity components are included, the inferred redshift-dependence is different, i.e. the posterior samples curve back towards $0$ in this fit, whereas they were monotonic in the previous fit, and the fit, while acceptable, is qualitatively worse, notably in times near mergers.}
    \label{fig:turbfits}
\end{figure}

Given some model-free indication that accretion, and gas flows more generally, can provide a measurable contribution to the gas velocity dispersion, we return to our explicit model (Eq. \ref{eq:KE}), and fit another version of it. In the first version we fit the total kinetic energy, but now we fit only to the velocity associated with random motion, i.e. $\mathrm{KE}_\mathrm{turb}(r,t) = N\cdot (\sigma_r^2 + \sigma_\phi^2 + \sigma_z^2)$, but leave the right-hand side unchanged. In other words, each term in the model is the same, with the same priors on the coefficients, but the parameters are now tuned to fit $\mathrm{KE}_\mathrm{turb}(r,t)$ instead of $\mathrm{KE}(r,t)$. In this fit there is now a large reservoir of energy, particularly the energy associated with $N \cdot v_\phi^2$, that may in principle be exchanged with $\mathrm{KE}_\mathrm{turb}(r,t)$, meaning that both $\dot{E}_\mathrm{visc}$ and $\dot{E}_\mathrm{diss}$ may represent physically different processes in this fit. Instead of just the generation of kinetic energy from (presumably mostly) gravitational forces and turbulent dissipation, respectively, they may now include any conversion of coherent motion (i.e. energy associated with $v_r$, $v_\phi$, and $v_z$), into or away from kinetic energy associated with the velocity dispersions. This interchange may occur as a result of nonaxisymmetric motions, or even uneven transport within the disk that may affect the full distribution of velocities in any one annulus, affecting both the mean and variance of that component.

The results of one of these fits, in the same radial annulus as shown earlier in Figure \ref{fig:singlecolumn} is shown here in Figure \ref{fig:turbfits}. Since the same right hand side of Equation \ref{eq:KE} is used, the basis functions, i.e. the dotted lines, i.e. each component assuming $\beta_i=1$, are the same as before, as is the region excluded from the fit above redshift $4.5$. The scale on the y-axes is about an order of magnitude smaller, since only the turbulent energy is being fit here, as opposed to the total kinetic energy. The redshift-dependence is also substantially different -- while each basis function is monotonic in time (since it is the cumulative integral over time of a positive- or negative-definite quantity, i.e. the total energy contributed by each process), in this fit the inferred contribution of most components first increases as before, but then decreases as $t \rightarrow 13.7$ Gyr. This is the result of the $(1+z)^{\alpha_z}$ factor in Equation \ref{eq:KE} with $\alpha_z>0.$ 

Perhaps the most consequential difference between the turbulent vs. full KE fit is that there are noticeable differences between the pink and the black lines in Figure \ref{fig:turbfits}, i.e. even though the fit is still acceptable, it is undoubtedly worse. In particular, the fit near $t= 8$ Gyr misses by a large margin both before and after the spike in $E_\mathrm{turb}$ recorded around that time (the result of a merger). Intuitively, this fit is more ``difficult'' for several reasons. First, the fit, as we have constructed it, always has the option of picking some nonzero initial value of $\mathrm{KE}_\mathrm{turb}$, pushing all of the values of $\beta_i \rightarrow 0$, and attributing all variation to the error term. This option is discouraged by the priors, which have $p(\beta_i) \rightarrow 0$ as $\beta_i \rightarrow 0$, but as we have seen, even when the model appears to be working well, the $\beta_i$ can occasionally be quite small. While this does not happen at this particular radius, the fits are particularly prone to this failure mode for $\mathrm{KE}_\mathrm{turb}$, which, unlike $\mathrm{KE}$, can find itself near zero for long periods of time. In this scenario, it is difficult for the model to exactly cancel all of the right terms to keep $\mathrm{KE}_\mathrm{turb} \ga 0$. Possible remedies include stronger priors on $\beta_i$, or a more flexible model, e.g. one where the different terms are allowed to have different redshift-dependencies.

\section{Discussion} \label{sec:disc}

\begin{figure}[h]
    \centering
    \includegraphics[width=1.0\textwidth]{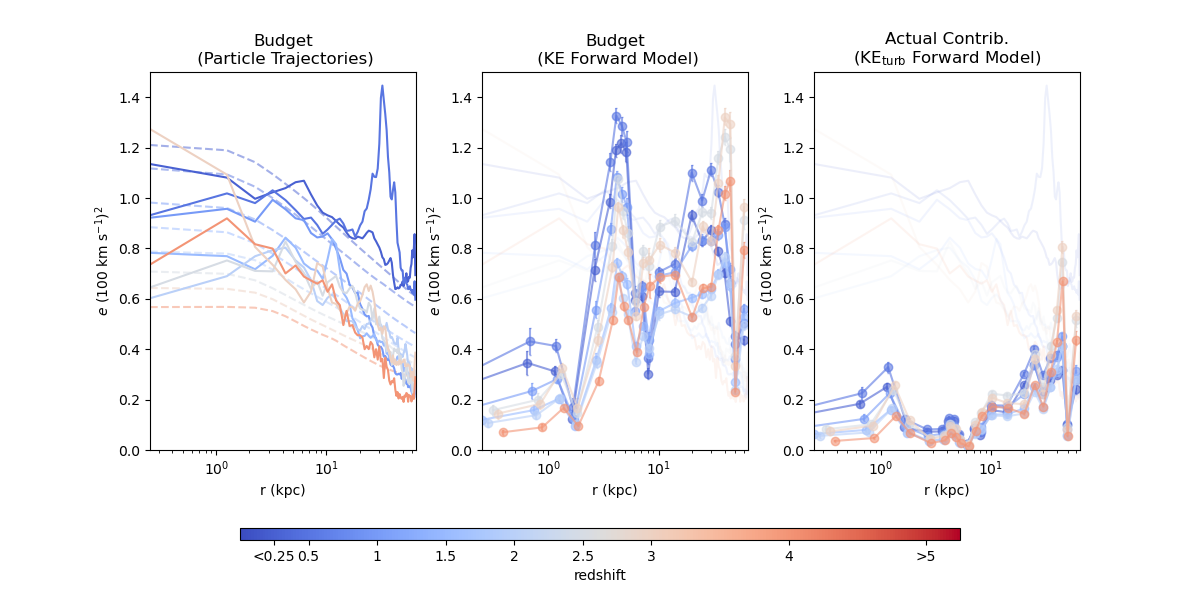}
    \caption{The specific energy imparted to the disk from accretion, measured three different ways as a function of radius. Lines from blue to red indicate different redshifts as in Figure \ref{fig:tracks}. The left-most panel shows the energies as calculated from the average energy lost by accreting tracer particles. These are compared to the dashed lines, showing mean values of $0.1 \Delta \Phi$, namely one tenth of the potential well depth at that radius. The middle and right panels show results from the explicit fits to KE and KE$_\mathrm{turb}$ respectively, and data from the left panel is lightly overplotted in the right two panels.. }
    \label{fig:specificEnergy}
\end{figure}

In the previous section, we made three different measurements of the specific energy associated with accreting tracer particles: the energy budget from the trajectories of individual tracer particles, and two forward models, one for the total kinetic energy, and the other for the turbulent kinetic energy. These three values are shown in Figure \ref{fig:specificEnergy} in bins of redshift and as a function of radius. We also showed model-free evidence of turbulence being driven by direct accretion by examining the auto-correlation functions of the time series $e_\mathrm{turb}$ and $\dot{M}_\mathrm{accr}$ at fixed radius. Our goal in this section is to interpret these results in the context of the literature, and prescribe plausible recipes for including these results in simplified models.

The specific energy associated with accretion events as measured both via the particle trajectories and the two explicit fits are shown in Figure \ref{fig:specificEnergy}. For the most part, the measurements from the particle tracks exceed those measured by the explicit models, and the contribution to the total kinetic energy exceeds the contribution to the turbulent kinetic energy. A reasonable interpretation is that the energy difference measured in the particle trajectories represents the maximum energy budget available, of order $(1/2) \dot{M} v_\mathrm{circ}^2$. The radial dependence of this energy difference appears to follow $\Delta \Phi$, the difference in the gravitational potential between the radius in question and the largest radius in our analysis (70 kpc). This suggests that the infalling tracer particles dissipate a large and fixed fraction of their free fall energy from the halo scale to the disk scale.

However, not all of this energy makes it to the disk itself. The forward model for the total kinetic energy should in principle be similar to the budget determined by the average particle trajectories if energy were conserved, i.e. all of the energy lost by the accreting particles were gained by the disk. This may be plausible in the outer part of the disk, given the overlap in the set of curves in the middle panel of Figure \ref{fig:specificEnergy} with the background curves reproduced from the first panel representing the particle-based energy budget, but in the inner part of the disk most of the energy is apparently lost. It is plausible that much of this energy is lost to radiation given that the inner disk has a much higher gas density.

Of the energy that makes it to the disk, presumably only some fraction will contribute to the driving of turbulence, measured in our case as the velocity dispersion in each annulus within the locally-measured scale height of the disk. In fact we do see that the contribution to the turbulent kinetic energy is generally strictly less than the contribution to the total energy, i.e. the curves in the rightmost panel of Figure \ref{fig:specificEnergy} lie below the curves in the center panel. There are many ways for the accretion energy to contribute to the total energy but not to the turbulent energy, including the driving of flows that change the average velocity of the $r$, $\phi$, or $z$ component of the velocity within the annulus. In principle some of this energy may be thought of as turbulent on a scale larger than the width of the annulus (and indeed much of the disk has a scale height greater than this width), while some of the ``turbulent'' energy may be large-scale flows in opposite directions on opposite sides of an annulus. These issues may be addressed with improved analyses, as in \citet{goldbaum_mass_2015}, although there is a trade-off in terms of the number of particles available for the analysis in the outer disk.

\begin{figure}[h]
    \centering
    \includegraphics[width=0.8\textwidth]{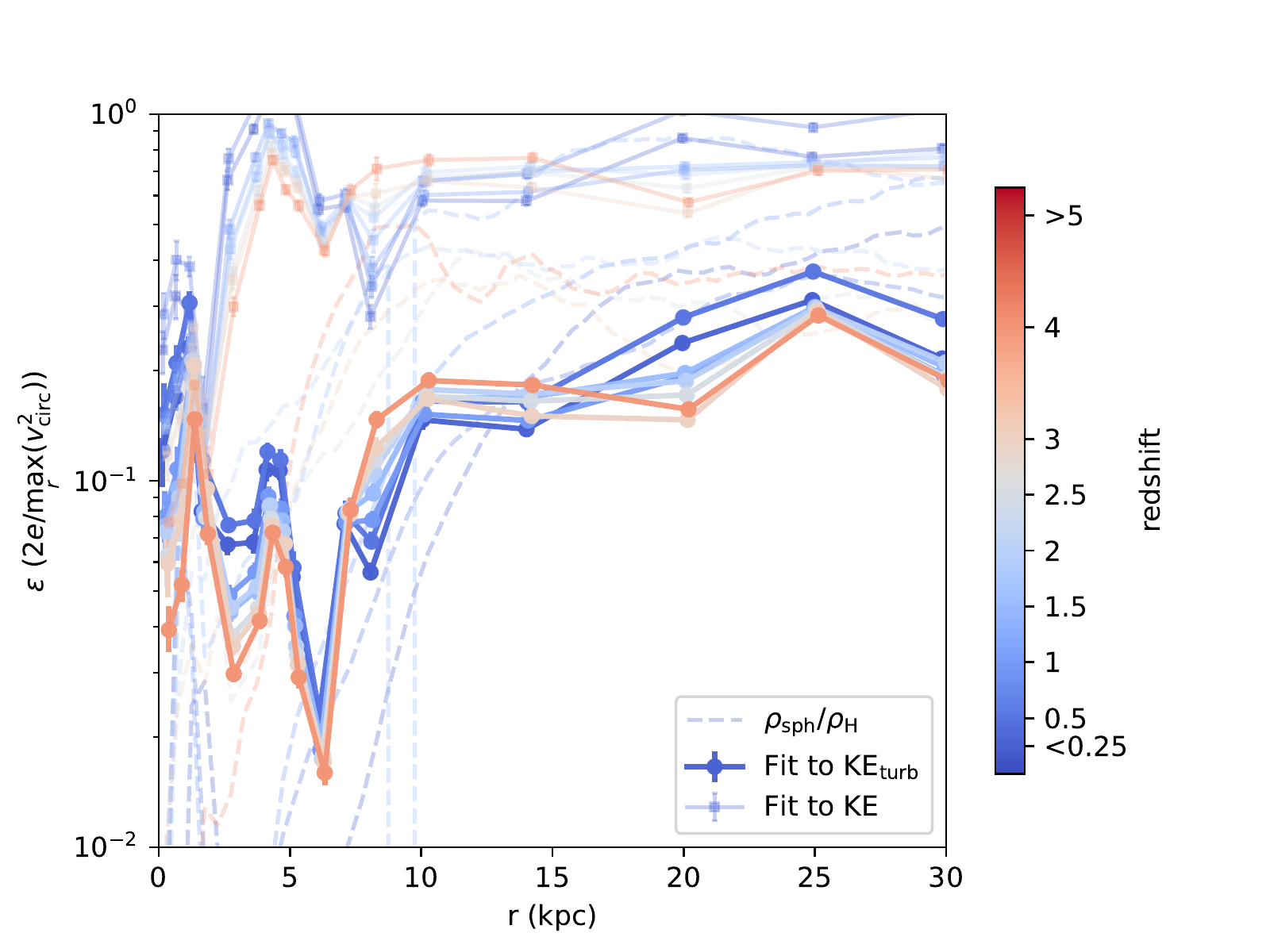}
    \caption{The efficiency of accretion at driving turbulence. The specific energy imparted by accretion as inferred by the forward models for KE and KE$_\mathrm{turb}$ are divided by the maximum circular velocity squared at each snapshot, averaged over the same redshift bins. The contribution of the accretion energy to KE$_\mathrm{turb}$ is smaller than the contribution to KE, likely reflecting energy losses at each stage. These curves are compared to simple estimates for the density contrast of the accreting material relative to the disk, which is the theoretical efficiency factor of \citet{klessen_accretiondriven_2010}, in the same redshift bins.}
    \label{fig:efficiency}
\end{figure}

\citet{klessen_accretiondriven_2010} found that the efficiency with which accretion drives turbulence, namely the fraction of the accretion energy per unit time $(1/2) \dot{M} v^2$ that drives turbulence, is roughly the density contrast between the accreting material and the object onto which the accretion is occurring. In Figure \ref{fig:efficiency}, we show the accretion efficiency as estimated by the explicit fits, i.e. the same quantities shown in the right two
panels of Figure \ref{fig:specificEnergy} divided by the maximum circular velocity squared at each timestep $\max_r v_\mathrm{circ}^2$. For comparison, we also show both the efficiency for the total kinetic energy, and estimates for the density contrast. The $\rho_\mathrm{sph}/\rho_H$ lines, corresponding to the gas density in a spherical shell of thickness 0.5 kpc divided by the gas density within a gas scaleheight, shares some qualitative features with the measured efficiencies, e.g. large values mostly independent of radius beyond $r\sim 15$ kpc, with lower values around $\sim 5$ kpc, and a rebound to larger values near $r \sim 0$ kpc, but quantitatively the agreement is not convincing.The measured efficiencies have an odd feature around 6 kpc where the fitted efficiencies dive by a factor of a few. While this feature may be an artifact of the imperfect fitting procedure, it may be at least partly driven by physical effects, including the growth of the star-forming disk (as can seen by the progressively larger radii at which the dashed lines transition from $\la 10^{-1}$ to $>10^{-1}$), and the change in geometry of the volume defining the disk around this radius (see Figure \ref{fig:rz}), which corresponds to a qualitative change in the particle trajectories from close to horizontal, to close to vertical.

Perhaps the simplest interpretation is that the efficiency with which accretion drives turbulence is of order 10 percent at all redshifts and radii, given that stronger redshift- or radial- dependence would be difficult to estimate a priori in a simplified model. If the model has access to a believable estimate of the density of extraplanar gas relative to the local gas density of the disk, some additional radial dependence may be warranted.

\subsection{Implications for Energy Balance in Disks}

We now turn to the question of where turbulent driving from accretion is important compared to other sources, namely flow of matter through the disk, and feedback from star formation. In what follows we will construct a simple flexible analytic framework under a minimalist set of assumptions to understand which source of turbulent energy sets the velocity dispersion. Essentially the {\it only} major assumption we make is that any location in the disk is in local energy balance
\begin{equation}
\label{eq:energybalance}
    \dot{E} = \dot{E}_\mathrm{SF} + \dot{E}_\mathrm{transport} + \dot{E}_\mathrm{accretion} - \dot{E}_\mathrm{diss} \approx 0\ \mathrm{for}\ \sigma>8\ \mathrm{km}\ \mathrm{s}^{-1},
\end{equation}
and that the accretion rate's dependence on radius follows a simple exponential form. Beyond that, one should think of this subsection as an exercise in visualization and plugging in reasonable numbers to the terms in this equation. In follow-up work, we will apply more constraints to generate a self-consistent data-driven prediction for the velocity dispersion profiles in disks. (For physical models with more assumptions, see \citet{forbes_radially_2019} and \citet{ginzburg_evolution_2022}). For each term in this equation, we adopt the same simple forms we have used in the introduction and throughout, namely per unit area in the disk, the energy contribution from star formation is $\dot{E}_\mathrm{SF} = \dot{\Sigma}_\mathrm{SF} \langle p_* / m_* \rangle \sigma$, while the contribution from gas moving through the disk is $\dot{E}_\mathrm{transport} = \dot{M} v_\mathrm{circ}^2 / (2\pi r^2)$, the contribution from accretion is $\dot{E}_\mathrm{accretion} = \epsilon \dot{\Sigma}_\mathrm{accr} v_\mathrm{circ}^2$, and turbulent dissipation is taken to be $\dot{E}_\mathrm{diss} = (3/2) \Sigma \sigma^3/H$.

\begin{figure}[h]
    \centering
    \includegraphics[width=0.9\textwidth]{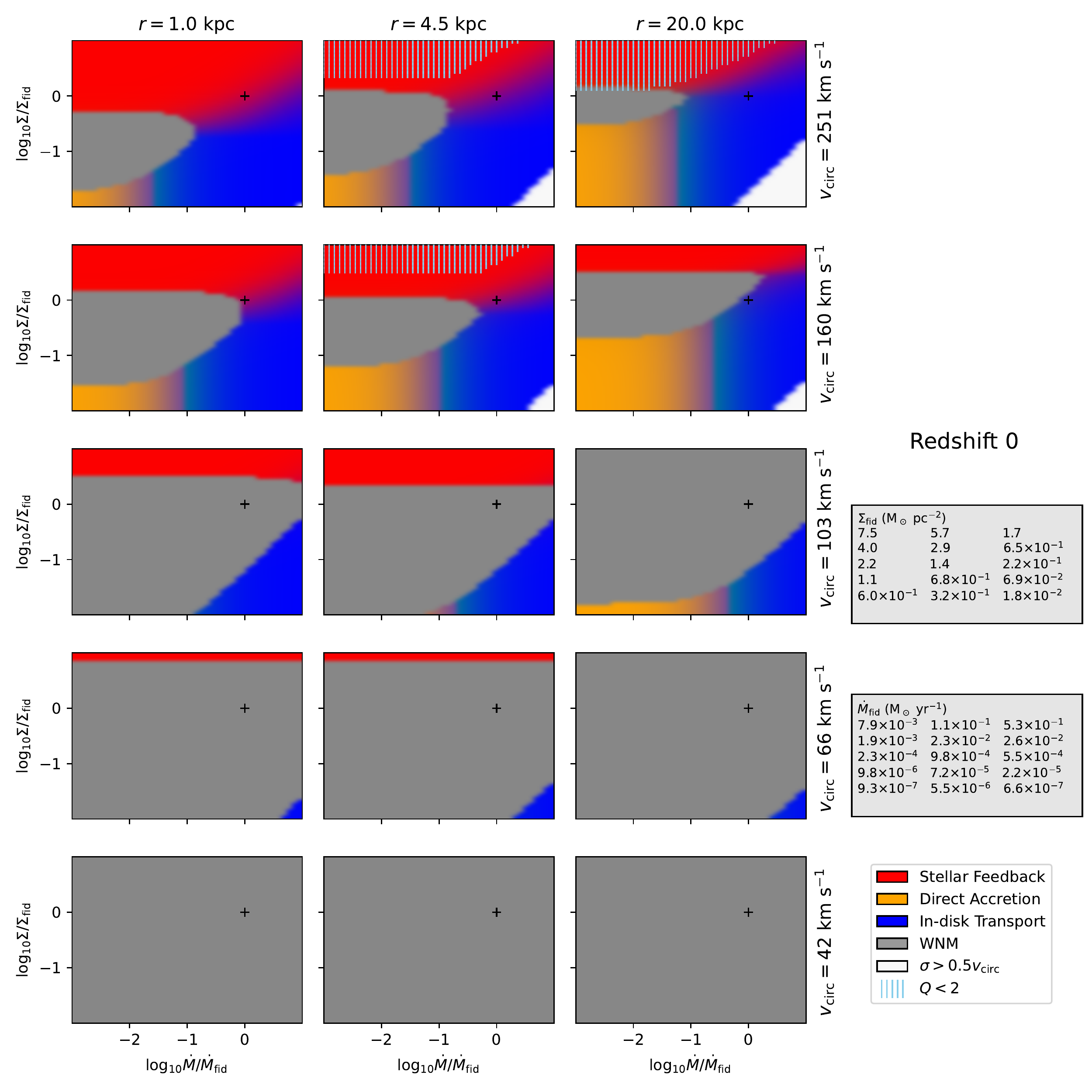}
    \caption{The relative importance of different sources of turbulent energy across parameter space at z=0. Red shows regions dominated by stellar feedback, blue shows in-disk inflow-dominated regimes, and orange shows where direct accretion dominates the energy budget. Gray shows regions where the velocity dispersion is below 8 km s$^{-1}$, which we take to mean that the velocity dispersion is set by the soundspeed of a warm neutral medium (WNM). In white regions, the velocity dispersion is greater than $v_\mathrm{circ}/2$, suggesting that it is inappropriate to consider that region a disk. Each column is a different galactocentric radius, and each row corresponds to a different galaxy circular velocity. In each panel, the column density and mass transport rate are varied systematically around fiducial values as specified in the text. These fiducial values are marked with a small $+$ at $(0,0)$ in each panel, and their values in physical units are shown as 3 by 5 tables above the legend, with each entry associated with its corresponding panel. Regions marked by vertical lines have $Q<2$, suggesting that disks are unlikely to remain in that state for very long. Lower-mass galaxies generally have low velocity dispersions arising from the WNM, while galaxies with circular velocities above $\sim 100$ km s$^{-1}$ may have turbulence set by any of the three source terms we have considered depending on the actual flow rate through these galaxies and their actual column density profiles. In general accretion and transport are more likely to be important at large radii.}
    \label{fig:regimes}
\end{figure}

\begin{figure}[h]
    \centering
    \includegraphics[width=0.9\textwidth]{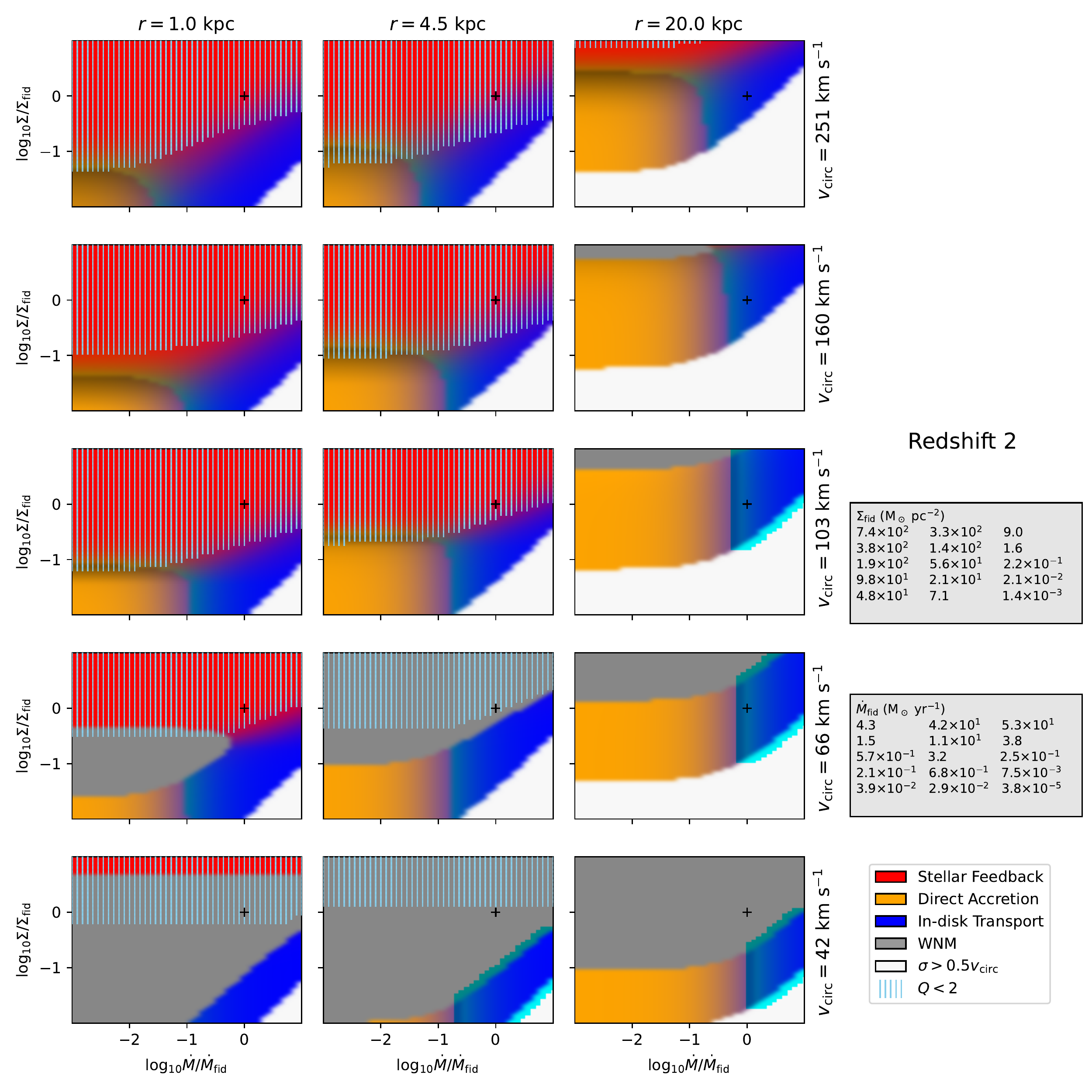}
    \caption{Same as Figure \ref{fig:regimes}, but for redshift 2 instead of redshift 0. Substantially more of the parameter space at $z\sim 2$ is excluded by requiring $Q \ga 2$ (regions that do not meet this requirement are marked by light-blue vertical lines). Galaxies over a wider range of mass may have velocity dispersions in excess of the WNM floor. Enough energy is being added in the outer regions of the disk that even modestly higher $\dot{M}$ or lower $\Sigma$ values would heat up the disk enough that its velocity dispersion would become comparable to the circular velocity (regions shown in white).}
    \label{fig:regimesz2}
\end{figure}

Assuming fixed values of $\epsilon$, which we fix at 10\% based on Figure \ref{fig:efficiency}, and $\langle p_*/m_* \rangle$ the relative importance of the 3 positive terms in Equation \ref{eq:energybalance} is set by 6 variables: $r$, $\sigma$, $\dot{\Sigma}_\mathrm{SF}$, $\dot{\Sigma}_\mathrm{accr}$, $v_\mathrm{circ}$, and $\dot{M}$. Assuming that the velocity dispersion $\sigma$ is in equilibrium as was shown to be generically the case in \citet{forbes_balance_2014}, a simple consequence of the equilibration time of the energy equation tending to be short, we can solve Equation \ref{eq:energybalance} for $\sigma$, so long as we also specify $\Sigma$ and estimate $H$. While it seems like we have exchanged 1 variable ($\sigma$) for two ($\Sigma$ and $H$) that we must plug in, we can now eliminate another variable, $\dot{\Sigma}_\mathrm{SF}$ by employing the \citet{krumholz_star_2013} H$_2$-regulated star formation law, provided that we can specify $Z$. The advantage of essentially exchanging $Z$ for these other possible variables we could specify is that, over the range of galaxies we are interested in, $Z$ varies only by about an order of magnitude and is easy to specify via the mass-metallicity relation once we know almost anything else about the galaxy.

We therefore begin to construct our visualization (Figures \ref{fig:regimes} and \ref{fig:regimesz2}) as follows: we pick a logarithmically-spaced range of $r$ and $v_\mathrm{circ}$ values, each combination of which will get its own panel. Within each panel, we pick a range of $\dot{M}$ and $\Sigma$. To solve Equation \ref{eq:energybalance}, the only things left to determine are $H$, $Z$, and $\dot{\Sigma}_\mathrm{accr}$. We also have to choose a reasonable range of $\Sigma$ and $\dot{M}$ to plot, and this range may in fact vary substantially based on the quantities that are already fixed ($r$ and $v_\mathrm{circ}$). Each panel therefore shows a range of $\dot{M}$ and $\Sigma$ relative to some fiducial value $\dot{M}_\mathrm{fid}$ and $\Sigma_\mathrm{fid}$ that we will specify below. 

Once we have fixed $v_\mathrm{circ}$, we can adopt a corresponding $M_*$ based on the observed Tully-Fisher relation from \citet{miller_assembly_2012}, which is more or less independent of redshift. This $M_*$ will be useful in several ways: first we can use it to estimate a halo mass via the redshift-dependent stellar mass-halo mass relation of \citet{moster_galactic_2013}, and second we can estimate the gas-phase metallicity $Z$ from the redshift-dependent compilation of observational data on the stellar mass-gas metallicity relation used in \citet{hayward_how_2017}. This metallicity, along with a modest metallicity gradient of -0.03 dex kpc$^{-1}$ \citep[e.g.][]{zaritsky_ii_1994} now fully specifies the metallicity in every panel of the figures. To estimate $H$, we assume vertical hydrostatic equilibrium, yielding the following quadratic equation: $H = \sigma^2 (\pi G(\Sigma + f\Sigma_* + H \rho_\mathrm{DM}))^{-1}$. Here $f$ is a monotonic function of $\sigma/\sigma_*$, which we take to be a constant $0.2$ for simplicity, and $\rho_\mathrm{DM}$ is the local dark matter density, which assumes a relationship between halo mass, redshift, and the halo's structural parameters from \citet{dutton_cold_2014}. This requires us to estimate the stellar surface density, which we take to be the stellar mass (already estimated) distributed in an exponential profile with a scale length from \citet{lang_bulge_2014}'s powerlaw fit for the disk scale lengths of star-forming galaxies, and scaled as $(1+z)^{-1}$. At this point all that remains is to determine what we will use for $\dot{\Sigma}_\mathrm{accr}$ and the fiducial values $\dot{M}_\mathrm{fid}$ and $\Sigma_\mathrm{fid}$ in each panel.

Following e.g. \citet{popping_evolution_2014}, we set the fiducial gas surface density in each panel to follow an exponential scale length 1.6 times larger than what we assumed for the stellar scale lengths, based on the observed ratio in local galaxies \citep{bigiel_universal_2012}, though note that there are good reasons to expect this profile not to apply at higher redshifts \citep[e.g.][]{krumholz_dynamics_2010, forbes_balance_2014}. The total mass in this fiducial profile is taken from the redshift-dependent dust-based fit of \citet{genzel_combined_2015}. The fiducial value of $\dot{M}$ is less straightforward. Several plausible options exist, but in general they may radically disagree.

The first option is to estimate $\dot{M}$ based on physical conditions in the disk at that location \citep[e.g.][]{dekel_formation_2009, krumholz_dynamics_2010, forbes_balance_2014, ginzburg_evolution_2022}. The difficulty here for our purposes is that these prescriptions will often entail large values of $\dot{M}$ that will radically rearrange the structure of the disk, meaning that whatever fiducial value we have adopted for the disk will not persist for very long. We therefore do not use this $\dot{M}$ for our fiducial estimate, but inspired by these models, we will check at the end of our procedure which areas of the parameter space would have $Q\la1$, and hence be subject to these large transport rates. The next option is to assert that the fiducial $\dot{M}$ will be such that the star formation + outflows interior to the radius in question can be exactly supplied by $\dot{M}$, in other words
\begin{equation}
 \dot{M}_\mathrm{supply-inner} = \int_0^r 2\pi r' \left( \dot{\Sigma}_\mathrm{SF}(r')(f_R + \eta) - \dot{\Sigma}_\mathrm{accr}(r') \right) dr'.   
\end{equation}
Here $\eta$ is the mass loading factor, the ratio of the outflow rate to the star formation rate, and $f_R \approx 0.45$ is the fraction of gas that is returned to the ISM as a result of stellar evolution \citep{tinsley_evolution_1980, leitner_fuel_2011}. This approach is similar to recent modeling by \citet{wang_origin_2022} and \citet{wang_gasphase_2022}. A final option is to make sure the inflow rate from cosmological accretion can be accomodated, i.e. 
\begin{equation}
 \dot{M}_\mathrm{accomodate-outer} = \int_r^\infty 2\pi r' \left( \dot{\Sigma}_\mathrm{accr}(r')  - \dot{\Sigma}_\mathrm{SF}(r')(f_R + \eta) \right) dr'.   
\end{equation}
The disadvantage of these two approaches is, first, that they assume the gas profile is close to equilibrium, and second, that they break the locality of our evaluation of the energy equation. Now what matters is not just $\Sigma/\Sigma_\mathrm{fid}$ in a given panel, but also the profile of $\Sigma$ and hence $\dot{\Sigma}_\mathrm{SF}$ at all interior or exterior radii. Moreover, these two values of $\dot{M}$ will not agree in general unless we also enforce global mass balance, namely that the total accretion rate equals the star formation plus mass outflow rate. The equilibrium assumption for the gas profiles is actually borne out in physical models \citep{forbes_balance_2014}, and global gas balance is generally a reasonable assumption, again because the equilibration time for the mass continuity equation tends to be short \citep{bouche_impact_2010, dave_analytic_2012, lilly_gas_2013, forbes_origin_2014}. We therefore adopt these latter estimates as $\dot{M}_\mathrm{fid}$ and guarantee that they are the same by enforcing global gas mass equilibrium (see next paragraph). We assume that the fiducial gas profile holds everywhere for the purposes of evaluating $\dot{M}_\mathrm{fid}$, which is not totally self-consistent with the range of $\Sigma \ne \Sigma_\mathrm{fid}$ shown in each panel, but at least fixes $\dot{M}_\mathrm{fid}$ to a reasonable value. For simplicity we adopt $\eta=1$ when evaluating $\dot{M}_\mathrm{fid}$.

All that remains now is to specify $\dot{\Sigma}_\mathrm{accr}$. Given this prescription for $\dot{M}_\mathrm{fid}$, the most consistent choice we can make for $\dot{\Sigma}_\mathrm{accr}$ is to set the total mass accreted to $\int_0^\infty 2\pi r' \dot{\Sigma}_\mathrm{SF}(r')(f_R + \eta) dr'$, where as before we evaluate $\dot{\Sigma}_\mathrm{SF}$ assuming the fiducial $\Sigma$ profile. This value will generally be less than other simple choices that could be made, e.g. assuming that the baryonic accretion is closely tied to the dark matter accretion rate times the cosmological baryon fraction \citep{bouche_impact_2010}. This reflects potential preventative feedback processes at work \citep[e.g.][]{okamoto_mass_2008, lu_formation_2015, pandya_first_2020}. Global equilibrium could also be enforced by increasing $\eta$, so by making the ``preventative'' feedback assumption, we may be underestimating the impact of accretion energy throughout the parameter space. The profile of the accretion mass surface density is assumed to be exponential with a scale length of 5\% of the Virial radius.

With all of these quantities set, $\sigma$ is determined numerically by finding the value for which energy injection is equal to energy dissipation. Given a solution, we can check which of the three sources of energy is dominant, which sets the color in that region of parameter space shown in Figure \ref{fig:regimes}. We set a floor of $\sigma = 8\ \mathrm{km}\ \mathrm{s}^{-1}$ on the solution to the energy balance equation under the assumption that the volume of the interstellar medium will typically be dominated by a phase similar to the Milky Way's warm neutral medium, which arises in almost all situations via UV heating from young stars \citep{bialy_thermal_2019}. Regions for which this floor on $\sigma$ has been enforced are colored gray in Figures \ref{fig:regimes} and \ref{fig:regimesz2}. 

Another special case that may arise is that the multi-component $Q$ summarizing the stability of the disk to gravitational perturbations may fall below $\sim 1$, indicating that the disk is unstable. In this situation, the disk is likely to rapidly adjust to a state of marginal stability \citep{bournaud_unstable_2009, goldbaum_mass_2015, goldbaum_mass_2016, behrendt_clusters_2016}. The critical value of $Q$ for marginal stability is likely not exactly 1, but modestly higher \citep{elmegreen_gravitational_2011}. For our purposes the simple \citet{wang_gravitational_1994} approximation to the multi-component $Q$, namely $1/Q \approx 1/Q_* + 1/Q_\mathrm{gas}$, suffices despite the availability of better estimates when more information about the state of the disk is available \citep{rafikov_local_2001, romeo_effective_2011}. To estimate $Q_*$ we assume $\sigma_* = 30\ \mathrm{km}\ \mathrm{s}^{-1}$ and that $Q_*$ itself must be at least 2.5 (see discussions in \citet{forbes_balance_2014}). In general $\sigma_*$ will not be constant across stellar populations \citep[e.g.][]{holmberg_genevacopenhagen_2009}, and will be the result of some combination of gradual heating due to gravitational encounters \citep{lacey_influence_1984, ida_origin_1993, donghia_selfperpetuating_2013} or large birth velocity dispersions \citep{bournaud_thick_2009, forbes_evolving_2012, birdOutUpsideDownRoles2020}. In Figures \ref{fig:regimes} and \ref{fig:regimesz2} we mark regions with $Q<2$ with vertical lines -- our expectation is that disks are unlikely to be found in these states except for brief transient phases, so just like regions with $\sigma$ at its floor value, the dominant source of turbulent energy in that regime is not particularly relevant. While the exact value of $Q$ that signifies marginal stability is uncertain \citep{elmegreen_gravitational_2011}, we note that in these diagrams the value of $Q$ changes rapidly across this $Q=2$ boundary, so these regions are not sensitive to the exact value of marginal stability. At $z\sim 2$, the fiducial values of $\dot{M}$ and $\Sigma$ end up with $Q < 2$, suggesting that these fiducial guesses, while seemingly reasonable, should be used with caution.

The general picture that emerges from Figures \ref{fig:regimes} and \ref{fig:regimesz2} is that stellar feedback is dominant in high column density regimes, but these regimes, especially at high redshift, often fall afoul of the expectation that the disk should not be violently gravitationally unstable, i.e. the expectation that $Q\ll 1$ disks will rapidly rearrange themselves to marginal stability. Smaller column densities and/or larger $\dot{M}$'s produce turbulence dominated by in-disk inflow, while lower $\dot{M}$'s, particularly at large radii, may be dominated by turbulence driven by direct accretion. While direct accretion-driven turbulence is certainly possible, it is not the norm unless the mass inflow rate and/or the column density are somewhat lower than their fiducial values.

Given this picture, it is reasonably clear why \citet{hopkins_accretion_2013} concluded that accretion does not drive turbulence in galaxies. First, their disks were set up to be unstable with $Q \approx 1$, meaning that much of their evolution is likely to be set by a rearrangement of the disk to regain marginal stability, as observed in e.g. \citet{goldbaum_mass_2015} and \citet{goldbaum_mass_2016}. Second, none of the configurations they tested fall into the region where we expect direct accretion to matter, except perhaps in the furthest outskirts of the disk. 

This brings up an important point about what is meant observationally by $\sigma$, and the quantities constructed by theorists to be comparable. Since the linewidth measurements on which $\sigma$ is based are often made in a line like H$\alpha$, generally associated with star formation, a single value of $\sigma$ for a given galaxy is weighted towards the values of $\sigma$ in the star-forming part of the disk. As a result theoretical comparisons are often constructed \citep[e.g.][]{forbes_radially_2019} as $\sigma \approx \int_0^\infty 2\pi r \sigma \dot{\Sigma}_\mathrm{SF} dr / \dot{M}_\mathrm{SF}$. These averages are often going to be dominated by velocity dispersions set by stellar feedback or in-disk flows, although this is not guaranteed in all regimes.

\section{Conclusion} \label{sec:conclusion}

The turbulence in galactic disks is a key intermediary in every physical process governing galaxies, including star formation, outflows, radial transport, and accretion. The velocity dispersion in the gas of galaxies, which includes and is often dominated by the turbulence, also has the advantage of being observable in galaxies at a variety of redshifts. Observationally, galaxies with higher star formation rates appear to have higher velocity dispersions. Much debate has arisen in the past few years about the physical origin of this correlation, particularly focusing on the influence of stellar feedback versus gas inflowing through the disk, thereby releasing gravitational potential energy.

In this work we used the highest resolution volume of the TNG suite of cosmological magnetohydrodynamic simulations, The TNG50 Simulation, the high time cadence output subboxes of the simulation, and the Lagrangian tracer particles to follow the motion of gas in and around one particular galaxy in incredible detail. The galaxy has a $z=0$ stellar mass of $5 \times 10^9 M_\odot$ and a circular velocity of 110 km s$^{-1}$, and our analysis employs data across the entire history of the simulation, though with a special emphasis on redshifts $<4.5$. We used four separate analyses of this galaxy to dissect the direct energy input of accretion on this galaxy:
\begin{enumerate}
    \item {\bf Stacking}: every tracer particle entering the disk volume is tracked backwards and forwards in time around when it enters the disk. We find that the tracers each lose a specific energy of $\sim (100\ \mathrm{km}\ \mathrm{s}^{-1})^2$, which may be shared with the disk. This level of specific energy is comparable to the circular velocity of the disk itself. The radial dependence of this average energy budget follows the halo potential well (see the left panel of Figure \ref{fig:specificEnergy}).
    \item {\bf Total KE forward model}: we assume every tracer particle entering or leaving any particular annulus in the gas disk is associated with a particular specific energy, {\it a priori} the circular velocity squared. The actual energy of each type of event is then fit with a nearly-linear model by requiring that all of the sources and sinks of energy add up to the kinetic energy in that annulus of the disk as a function of time. We find that accretion contributes modestly less to the energy of the annuli than the stacking analysis, which is consistent with some energy loss in the process of transferring the accreting particles' energies to the disk (see the center panel of Figure \ref{fig:specificEnergy}).
    \item{\bf Turbulent KE forward model}: Rather than fitting the full kinetic energy in each annulus, we fit just the part of the kinetic energy associated with random motions in this annulus, i.e. the mass times $\sigma_r^2 + \sigma_\phi^2 + \sigma_z^2$. We find that accretion drives random motion in the disk at a level of $\sim 0.1 (100\ \mathrm{km}\ \mathrm{s}^{-1})^2$, implying an efficiency of around 10\% (see Figure \ref{fig:efficiency}).
    \item {\bf Cross-correlations}: We cross-correlate the time series of the specific kinetic energy associated with random motions in the disk with other time series in each annulus, to look for significant correlations. We find asymmetries around a lag-time of zero for two of these: the accretion rate and the rate of inward mass flow in the disk, suggesting model-independent causal relationships between these quantities and the turbulent kinetic energy (see Figure \ref{fig:crosscorr2d}).
\end{enumerate} 
\noindent
The high time cadence, the resolution of the simulation, and the Lagrangian tracer particles allow us to make these novel measurements, following the mass and energy budget in every part of this galaxy in unprecedented detail.

We find that the turbulent energy budget in this TNG50 galaxy is dominated by radial flows in the interior (see Figures \ref{fig:singlecolumn} and \ref{fig:turbfits} which show the fits in the annulus at $r=3.75 \pm 0.25$ kpc) and accretion in the extreme outskirts. However, interestingly, the rate of turbulent dissipation appears to be much lower than expected (see Figures \ref{fig:singlecolumn}, \ref{fig:turbfits}, and the ``Dissipation'' panel of Figure \ref{fig:efac}) -- at this stage it is unclear whether this is a shortcoming in the fitting procedure, or a consequence of the numerical resolution, which does not resolve the scale height of the disk with a particularly large number of resolution elements. In the former case, our conclusions about the efficiency of accretion will be suspect, while in the latter case, it is possible that TNG's velocity dispersions will be overestimated owing to a relative paucity of dissipation. We have some evidence for the latter case, namely the time-asymmetric and model-free cross-correlation between the specific energy of the gas and the accretion rate, which strongly suggests a causal relationship between the two. We also note that the fits for the specific energy associated with each sort of event type are largely consistent across the many independent fits we conducted at different radii, suggesting that if the fits are unreliable they are somehow failing in the same way, which is possible but unlikely.

Proceeding under the assumption that the fits are reliable, we find that turbulence in the galaxy is driven by accretion with an efficiency of order 10\% relative to $(1/2) \dot{M}_\mathrm{accr} v_\mathrm{circ}^2$. As a result, we find that direct accretion is likely to be a substantial driver of turbulence in galaxies in the intermediate mass regime (circular velocities from $\sim 100-200$ km s${-1}$), at large radii ($\ga 10$ kpc), and/or for galaxies with relatively low gas mass surface densities or mass transport rates. This contribution to the turbulent energy is straightforward to include in simplified models, which could be used to explore the observational correlation between galaxies' star formation rates and velocity dispersions. As discussed in the introduction, this correlation may be a causal relationship or one where variations in both quantities are driven by a common cause, e.g. high surface densities simulatneously driving gravitational instability and high star formation rates, or high accretion rates simultaneously driving turbulence and increases in the surface density. We also expect that this methodology, namely using Lagrangian tracer particles in high time-cadence snapshots of cosmological simulations, may be helpful in understanding other phenomena in the simulations including AGN feedback, mergers, radial mixing and flows, and the formation and effect of bars.

\begin{acknowledgments}
We thank the IllustrisTNG team and the authors of the open source packages used in this study for making their simulation data and code publicly available. We thank Dan Foreman-Mackey, Du\v{s}an Kere\v{s}, Phil Hopkins, Robyn Sanderson, Alex Gurvich, and Julianne Dalcanton for helpful conversations. JCF is supported by a Flatiron Research Fellowship through the Flatiron Institute, a division of the Simons Foundation.  RE acknowledges the support by the Institute for Theory and Computation at the Center for Astrophysics. DN acknowledges funding from the Deutsche Forschungsgemeinschaft (DFG) through an Emmy Noether Research Group (grant number NE 2441/1-1. MRK acknowledges support from the Australian Research Council through its Future Fellowships funding scheme, award FT180100375. GLB acknowledges support from the NSF (OAC-1835509, AST-2108470), a NASA TCAN award, and the Simons Foundation.
\end{acknowledgments}

\software{numpy \citep{vanderwalt_numpy_2011,harris_array_2020},
          matplotlib \citep{hunter_matplotlib_2007},
          scipy \citep{virtanen_scipy_2020},
          emcee \citep{foreman-mackey_emcee_2013,foreman-mackey_emcee_2019}
          schwimmbad (\url{https://schwimmbad.readthedocs.io/en/latest/}),
          arrow (\url{https://arrow.apache.org/docs/index.html}),
          parquet (\url{https://parquet.apache.org/},
          mpi4py \citep{dalcin_mpi_2005, dalcin_mpi_2008, dalcin_parallel_2011, dalcin_mpi4py_2021}
          }

\appendix

\section{Galaxy centering and orientation }

\label{app:smoothing}

\begin{figure}[h]
    \centering
    \includegraphics[width=0.8\textwidth]{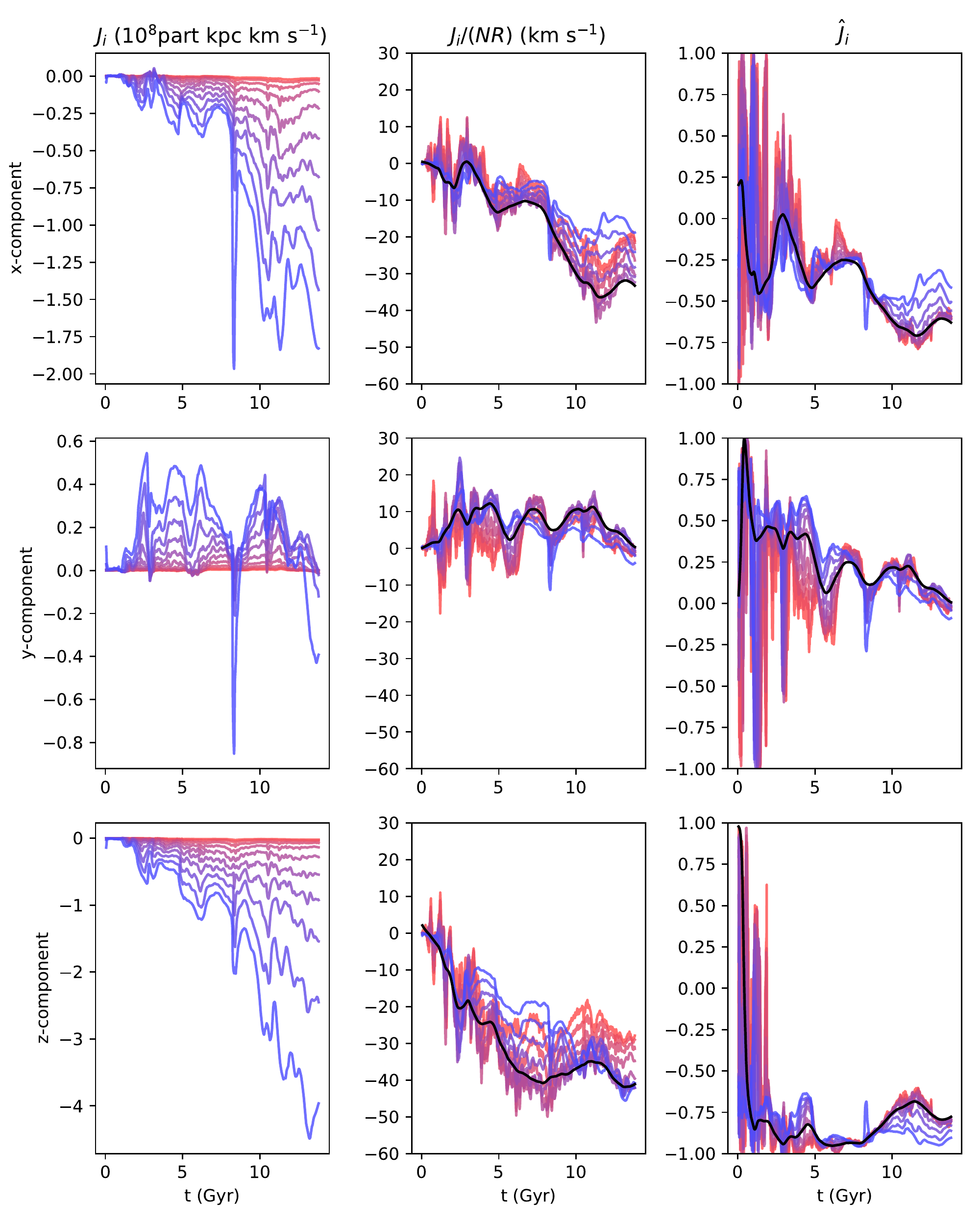}
    \caption{Angular momentum components as a function of time. Different lines from red to blue show larger radius spheres in which the angular momentum is computed, logarithmically spaced from $10^{0.4}$ to $10^{1.5}$ kpc. The black line shows the smoothed values of the angular momentum adopted as the galaxy's orientation throughout this work.}
    \label{fig:smoothJ}
\end{figure}

Much of our analysis in this work depends on careful accounting of the trajectories of the Lagrangian tracer particles. Of interest are both their velocities as a function of time, and their positions, which determine whether they are in a given radial bin or not, and whether they have undergone an accretion, outflow, or other event in between adjacent snapshots or not. It is therefore critical to choose a reference frame for the galaxy's cylindrical coordinate system, i.e. a position, velocity, and orientation, that does not change dramatically on short timescales.

We proceed iteratively: at each snapshot we identify the most-bound baryonic particle with which any of the tracer particles in our sample is associated. It turns out that at high redshift this most-bound particle moves around dramatically, i.e. among different progenitor halos. We check for this by examining the consistency of the velocity of the galaxy with the galaxy's change in position, and exclude candidate galaxy centers that are not close enough to the path of the galaxy's center, which is continuous to $z=0$, given its velocity. In particular we require that $|\Delta {\bf x}| < 2 |\Delta {\bf v}| \Delta t$. Typically these candidate galaxy centers are many kpc away from this acceptable range, so they are excluded by iteratively making cuts on acceptable $x$, $y$, and $z$ values in the full box coordinate system where the center is not allowed to be.

To avoid uncertainty on the scale of the baryonic resolution in the galaxy's center, we compute the center of mass of the baryonic material in spherical regions, first around the most-bound particle at a given snapshot, then around the center of mass. This procedure rapidly converges. The velocity is then taken to be the mean baryonic velocity in the same spherical region. This procedure is insensitive to exactly what size sphere we use -- in practice we adopt 10 pkpc.

The choice of orientation for the galaxy is much less obvious. Physically since angular momentum may easily be dominated by material at larger radii, there is no obvious ``right'' radius to choose. Moreover, the angular momentum is subject to short timescale variations at any radius, presumably as the result of substructures in the gas and stars entering or even leaving a given spherical radius. We therefore employ an explicit smoothing model, which itself requires a choice in parameters.

In particular, we fit a damped random walk model for each component of the angular momentum normalized by the mass and radius within a given spherical volume. This quantity, while obviously not conserved, at least removes much of the trend in the angular momentum itself from the growing mass of the galaxy. Each smoothed time series may then be combined together into a unit vector with no danger of unphysical values (i.e. unit vector components with magnitude greater than 1). Denoting the quantity to be smoothed $y$ (in this case $y = J_i/(N R)$, the $i$th dimension of the angular momentum vector with each dimension fit separately), we assume
\begin{equation}
    z_j = z_{j-1} + \epsilon_j
\end{equation}
where $j$ indexes the timestep, and the $\epsilon_j$ are assumed to be normally distributed with some mean $\mu$ and variance $(\Delta t_j \sigma_\mathrm{\Delta t})^2$. Here $\Delta t_j$ are the time differences between adjacent snapshots, and $\sigma_{\Delta t}$ is a parameter that we vary quantifying the allowed variation of the $z$ per unit time. To fully specify the model, we define a gaussian likelhood so that 
\begin{equation}
    \tilde{y}_j | z_j, \sigma \sim \mathcal{N}(z_j, \sigma^2) .
\end{equation}
The $\tilde{y}$ are a transformed version of $y$ -- an ordinary least squares quadratic fit for $y(t)$ is subtracted off so that $\tilde{y} = y - (\beta_0 + \beta_1 t + \beta_2 t^2)$ to remove any long-term trends before the gaussian random walk is fit. Finally we set the prior on $\sigma$ to be half-normal (if $w$ is normally-distributed about zero then $|w|$ is distributed as a half-normal), with width 3 km s$^{-1}$, and an uninformative prior on $\mu$, namely a normal distribution with width 1000 km s$^{-1}$.

The result is shown as the black lines in Figure \ref{fig:smoothJ}, where we have chosen $\sigma_{\Delta t} = 6.95$ km s$^{-1}$ Gyr$^{-1}$ and a spherical volume of radius 10.26 kpc. This choice results in a choice of orientation for the disk (the final column in Figure \ref{fig:smoothJ}) that closely follows the orientation of the baryons for any choice of spherical volume below 10 kpc, all of which agree with each other at low redshift.

\bibliography{tngAccretion}
\bibliographystyle{aasjournal}

\end{document}